\documentclass[prb,superscriptaddress]{revtex4-1}

\usepackage{amsmath}
\usepackage{amsfonts}
\usepackage{amssymb}
\usepackage{graphicx}

\begin{document}

\title{VO$_2$ nanosheets: controlling the THz properties through strain engineering}
\author{Elsa Abreu$^\S$}
\affiliation{Department of Physics, Boston University, Boston MA
02215, USA}
\author{Mengkun Liu$^\S$}
\affiliation{Department of Physics, Boston University, Boston MA
02215, USA}
\author{Jiwei Lu$^\S$}
\affiliation{Department of Materials Science and Engineering, University of Virginia, Charlottesville VA 22904, USA}
\author{Kevin G. West}
\affiliation{Department of Materials Science and Engineering, University of Virginia, Charlottesville VA 22904, USA}
\author{Salinporn Kittiwatanakul}
\affiliation{Department of Physics, University of Virginia, Charlottesville VA 22904, USA}
\author{Wenjing Yin}
\affiliation{Department of Materials Science and Engineering, University of Virginia, Charlottesville VA 22904, USA}
\author{Stuart A. Wolf}
\affiliation{Department of Physics, University of Virginia, Charlottesville VA 22904, USA}
\affiliation{Department of Materials Science and Engineering, University of Virginia, Charlottesville VA 22904, USA}
\author{Richard D. Averitt}
\affiliation{Department of Physics, Boston University, Boston MA 02215, USA}
\email{raveritt@physics.bu.edu}

\begin{abstract}

We investigate far-infrared properties of strain engineered vanadium dioxide nanosheets through epitaxial growth on a (100)$_R$ TiO$_{2}$ substrate. The nanosheets exhibit large uniaxial strain leading to highly uniform and oriented cracks along the rutile c-axis. Dramatic anisotropy arises for both the metal-insulator transition temperature, which is different from the structural transition temperature along the c$_R$ axis, and the metallic state conductivity. Detailed analysis reveals a Mott-Hubbard like behavior along the rutile c$_R$ axis.

$^\S$contributed equally to this work

\end{abstract}

\maketitle

\begin{center}
\textbf{I. INTRODUCTION}
\end{center}

During the past five decades, vanadates have been intensely investigated from the point of view of clarifying the physics of metal-insulator transitions (MIT). Vanadium dioxide (VO$_2$) is particularly intriguing, exhibiting a MIT with a conductivity decrease of over five orders of magnitude \cite{Ladd1969}. This is accompanied by a transition from a high temperature metallic rutile structure to an insulating monoclinic structure below the transition temperature. In bulk, the structural transition temperature, $T_{St}^{bulk}$, equals the metal-insulator transition temperature, $T_{MI}^{bulk}$, i.e. $T_{St}^{bulk}\simeq T_{MI}^{bulk} \simeq 340K$ \cite{Andersson1956}. However, VO$_2$ is not yet fully understood regarding the Mott-Hubbard or Peierls like nature of its MIT, though it is now generally accepted that a complete description requires explicit consideration of
electron-electron correlations \cite{Biermann2005, Lazarovits2010}. This is motivated in particular by the observed difference between the MIT and the structural transition temperatures, which provides significant evidence that the correlated conductivity behavior is independent of the structural phase \cite{Kim2006, Qazilbash2011, Arcangeletti2007}. Additionally, recent experimental studies highlight the multifunctional potential of VO$_{2}$ including current or photoinduced phase changes (with the commensurate large tuning of the dielectric function) as well as integration with other technologies for applications including light detection and memory-based metamaterials \cite{Kim2005, Driscoll2009,Hilton2007}.

Strain engineering enables an additional degree of control of technologically relevant properties and provides a discriminatory capability towards obtaining fundamental insight into the microscopic origin of the macroscopic characteristics. Strain has been used to modify the MIT temperature in VO$_2$ through direct application \cite{Muraoka2002} or substrate-dependent application \cite{Lu2008, Zhang2009} of stress. For the present measurements, we utilize highly strained epitaxial (100)$_R$ VO$_2$ thin films. The far-infrared conductivity is measured using non-contact polarization sensitive terahertz time-domain spectroscopy (THz-TDS) providing direct access to the coherent quasiparticle response along b$_R$ or c$_R$ by simply changing the sample orientation. Subsequent analysis indicates that the strain engineered tuning of vanadium $a_{1g}$ and $e_g^\pi$ orbitals controls the MIT transition temperature along c$_R$, which is different from the structural transition temperature. The origin of the observed conductivity anisotropy is also discussed, along with the potential for a technological application of strain engineered VO$_2$ thin films as temperature switched far-infrared polarizers.

\begin{center}
\textbf{II. EXPERIMENTS}
\end{center}

The $\sim100nm$ and $\sim250nm$ thick VO$_2$ films we investigated were deposited on rutile (100)$_R$ TiO$_2$ substrates by temperature optimized sputtering from a vanadium target, using the reactive bias target ion beam deposition technique in an Ar+O$_2$ gas mixture; details of the growth conditions can be found elsewhere \cite{West2008}. The samples morphology was characterized by optical, atomic force (AFM) and scanning electron microscopies (SEM), and the film microsctructure was analyzed by temperature dependent X-ray diffraction (XRD). 

THz-TDS is a non-contact method to measure far-infrared conductivity. The conductivity anisotropy is easily determined from transmission of THz pulses for different sample orientations.
In our case the output of a 1kHz 35fs Ti:Sapphire amplifier is used to generate nearly single-cycle THz pulses via optical rectification in a ZnTe crystal. We then employ a standard THz Time Domain Spectroscopy (THz TDS) setup to measure the transmission in the THz frequency range $\sim 0.2 - 2.0$THz \cite{Jepsen2011}. By changing the sample orientation with respect to the THz pulse polarization we are able to monitor the transmission along different crystal axes.

\begin{figure} [ptb]
\begin{center}
\includegraphics[width=0.5\textwidth,keepaspectratio=true]{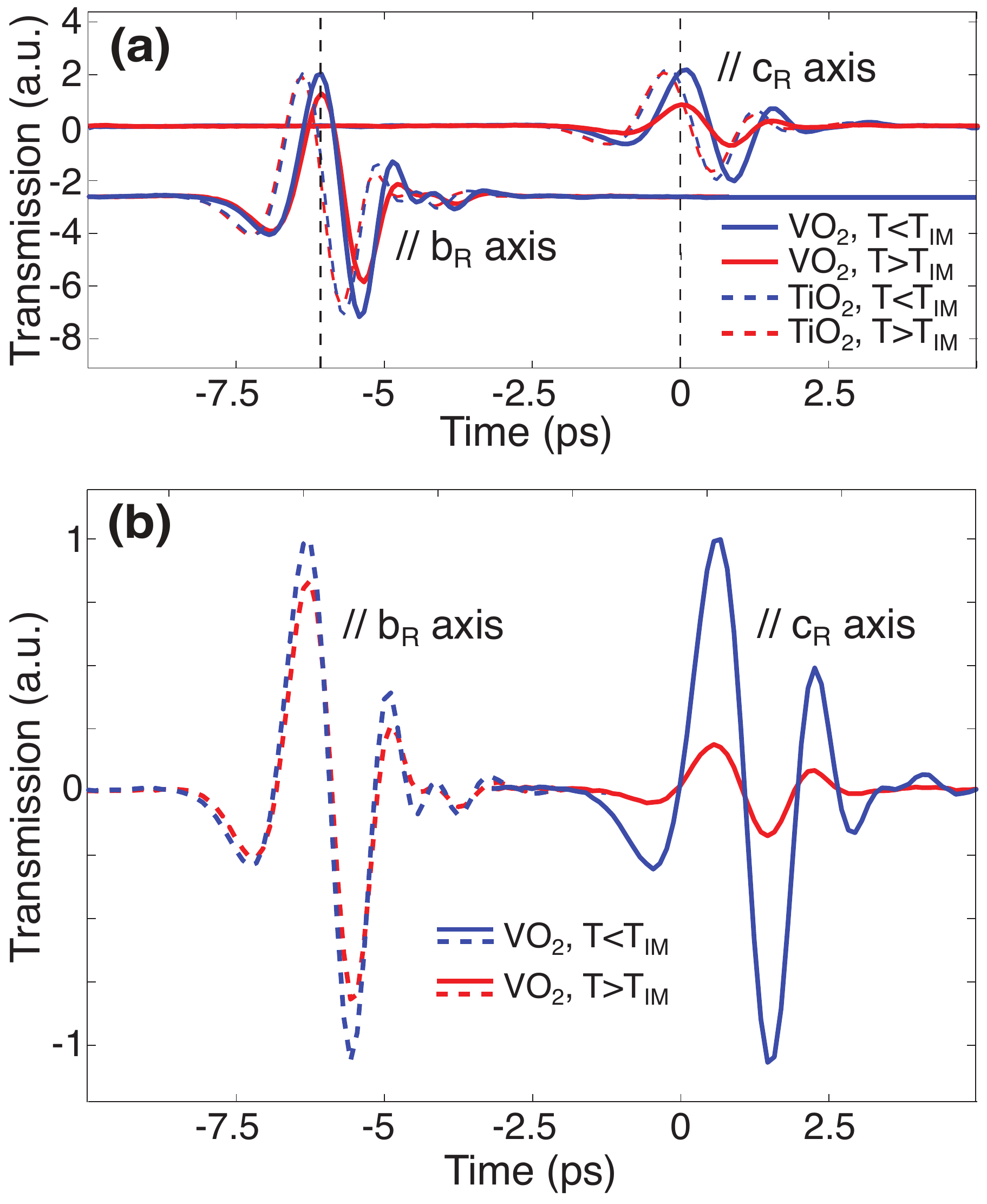}
\caption{(a) THz transmission along c$_R$ and b$_R$ (vertically offset) in the $100nm$ (100)$_R$ VO$_2$ sample (solid) and in the TiO$_2$ reference substrate (dashed), below (blue) and above (red) T$_{MI}$. The large refractive index anisotropy of the TiO$_2$ substrate leads to a slower propagation of the laser pulse along c$_R$ than along b$_R$, which enables the distinction of the two signals, and thus the orientation of the sample with respect to the incident field polarization. (b) THz transmission along c$_R$ and b$_R$ in the $250nm$ (100)$_R$ VO$_2$ sample, below (blue) and above (red) T$_{MI}$, normalized to the high temperature value. The relative low temperature transmission along b$_R$ ($\sim 85\%$) is dramatically different from that along c$_R$ ($\sim 15\%$).}
\label{waveForm}
\end{center}
\end{figure}

(100)$_R$ TiO$_2$ has a large refractive index anisotropy in the THz range, partly caused by its characteristic rutile structure. Such an anisotropy in the (100)$_R$ TiO$_2$ substrate makes it easy to distinguish transmitted THz signals with polarization parallel to c$_R$ from THz signals with polarization perpendicular to b$_R$ (Fig. \ref{waveForm}(a)).
Temperature dependent experimental characterization of the TiO$_2$ substrate in the THz range, using our THz TDS setup,
led to values of
\begin{eqnarray}
n_e\simeq 12.4-6\times10^{-4}\Delta T + 1.2i \nonumber \\
n_o\simeq9.1-3\times10^{-4}\Delta T + 0.4i \nonumber
\end{eqnarray}
for the refractive index along c$_R$ and b$_R$, respectively, in accordance with previous results \cite{Jordens2009}. Here,
$\Delta T$ stands for the temperature deviation with respect to room temperature. This approximation to the TiO$_2$ complex refractive index holds between room temperature and 400K; it does not take into account the temperature dependence of the imaginary part, which is negligible for our purposes.

The conductivity of the VO$_2$ film is extracted using the Fresnel equations, after experimental determination of the ratio of the THz transmission of the film to the THz transmission of a bare TiO$_2$ substrate, used as a reference.
THz TDS is thus a non-contact conductivity measurement, which allows one to quickly characterize the conductivity anisotropy in many samples with different thicknesses and substrates.

\begin{center}
\textbf{III. RESULTS}
\end{center}

\begin{center}
\textbf{A. Characterization of the VO$_2$ nanosheets}
\end{center}

\begin{figure} [ptb]
\begin{center}
\includegraphics[width=0.5\textwidth,keepaspectratio=true]{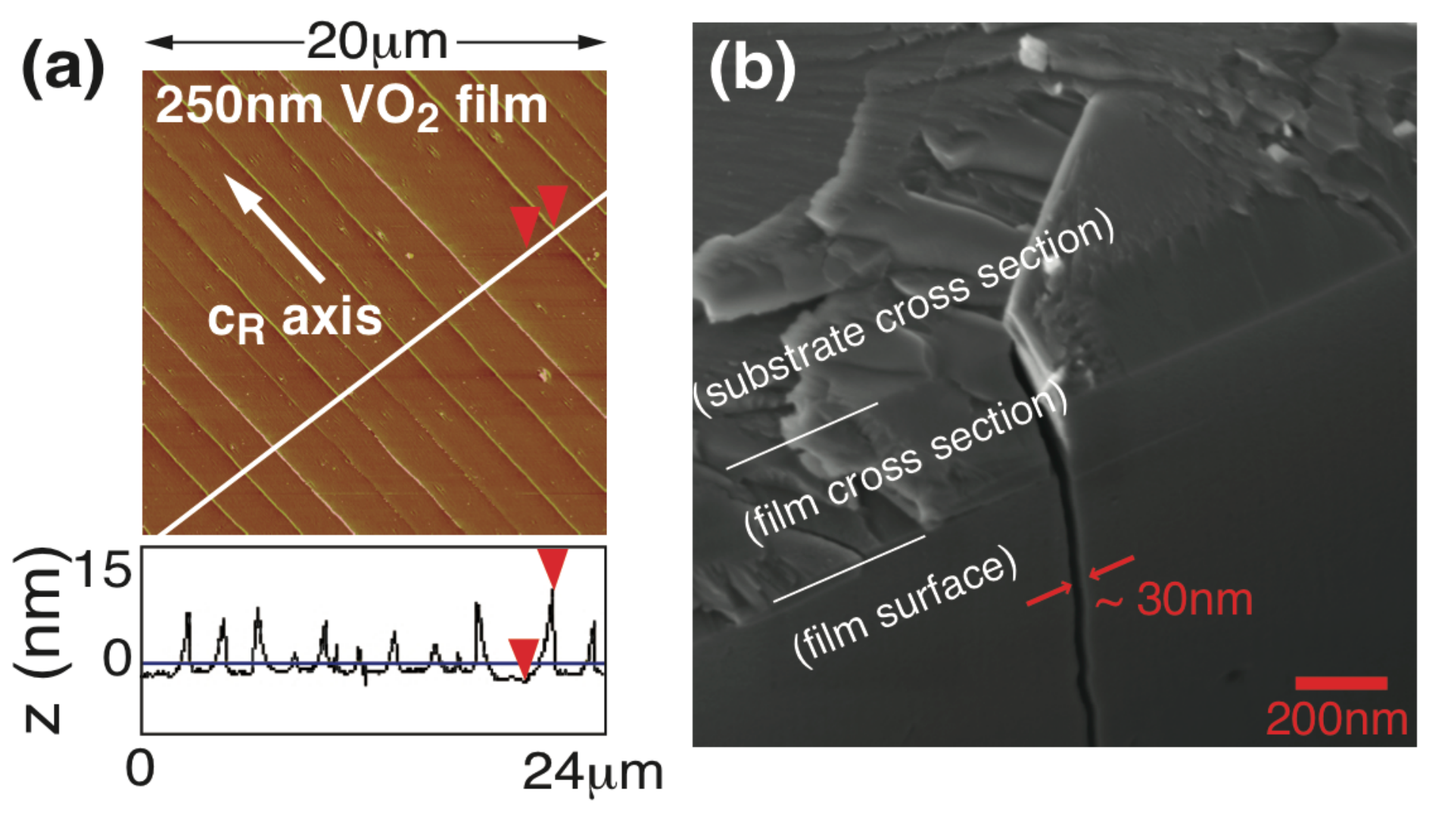}
\caption{Characterization of the $250nm$ thick (100)$_R$ VO$_2$ film. (a) AFM phase image (0$^\circ$-5$^\circ$ scale) and
corresponding height profile: the $250nm$ sample shows buckling  induced ridges along c$_R$ (height indicated by the arrows: $\Delta z=14.307nm$). (b) SEM image of a section of the sample (seen from the edge, the surface and the cross section being located as labeled) showing a $\sim 30nm$ wide crack.} \label{AFMSEM}
\end{center}
\end{figure}

Figure \ref{AFMSEM}(a) is an AFM phase image of the $250nm$ film, showing periodic buckling and cracking of the film parallel to c$_R$, with inter-crack spacings on the order of $1 \mu m$. The depth profile reveals nanosized ridges, $\sim 15nm$ high, near the cracks. The SEM image in Fig. \ref{AFMSEM}(b) confirms the $\sim 250nm$ thickness of the film. It shows that the cracks' depth matches the film thickness and it gives an estimate of about $30nm$ for their width. Such nanocracks were not detected by the AFM due to lack of tip sensitivity.

\begin{figure} [ptb]
\begin{center}
\includegraphics[width=0.5\textwidth,keepaspectratio=true]{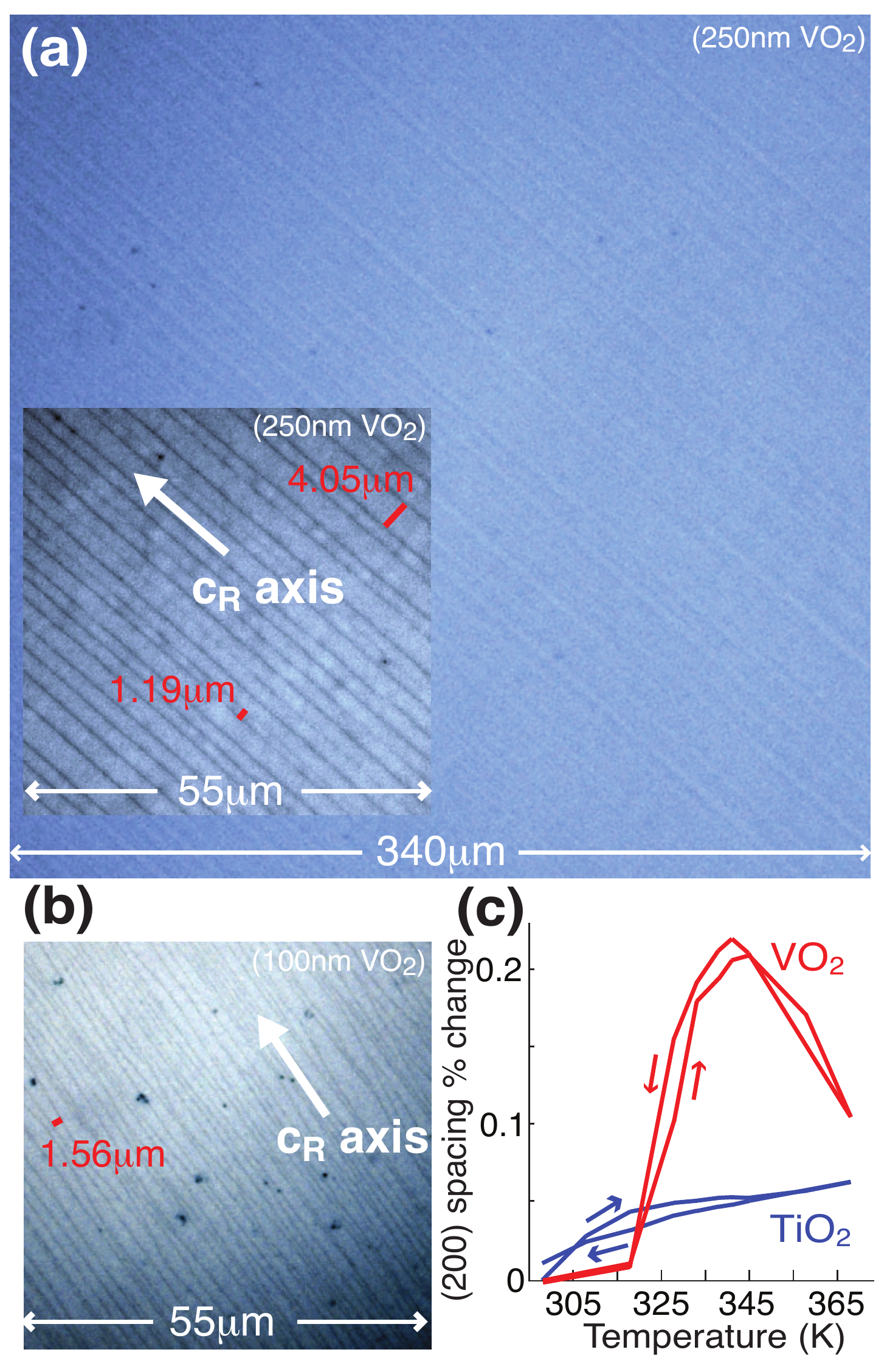}
\caption{Characterization of VO$_2$ thin films on a (100)$_R$ TiO$_2$ substrate. Optical images of the (a) $250nm$ and (b)
$100nm$ films: the samples show buckling induced ridges along c$_R$, spaced by $\sim 1\mu m$. (c) Temperature dependence (for increasing and decreasing temperature) of the a-axis lattice spacing, deduced from XRD data: a $\sim 0.1\%$ increase is observed along a$_R$, in the $100nm$ thick VO$_2$ sample, across the structural transition which occurs at $T_{St}=340K$.  The expected linear increase of the a$_R$-axis lattice spacing of TiO$_2$ with temperature is also observed.}
\label{optCharacterization}
\end{center}
\end{figure}

The optical images of the $100nm$ (Fig. \ref{optCharacterization}(a)) and the $250nm$ (Fig. \ref{optCharacterization}(b)) films confirm the $\sim 1\mu m$ period spacing of the cracks along c$_R$. Also, the observed uniformity in the distribution of cracks is an indication that our films are strained in a highly homogeneous and oriented fashion. This crack distribution enables a comparison of our results with those of VO$_2$ nanobeams, aligned along c$_R$ \cite{Cao2009, Zhang2009, Wu2006, Sohn2009, Jones2010}, although the dimensionality of our nanosheets gives access to the properties of strained VO$_2$ along more than one axis. Such cracking uniformity was achieved by optimizing the growth conditions of the samples, in particular the growth temperature. In the case of our (100)$_R$ VO$_2$ films the growth temperature was set to $\sim 500^\circ C$. Films grown at $\sim 450^\circ C$ show cracking along different directions, which makes the characterization of the films along c$_R$ less straightforward.
All subsequent analysis is analogous for both the $100nm$ and the $250nm$ thick films. Unless otherwise specified the results will refer to the $100nm$ thick sample.

\begin{figure} [ptb]
\begin{center}
\includegraphics[width=0.5\textwidth,keepaspectratio=true]{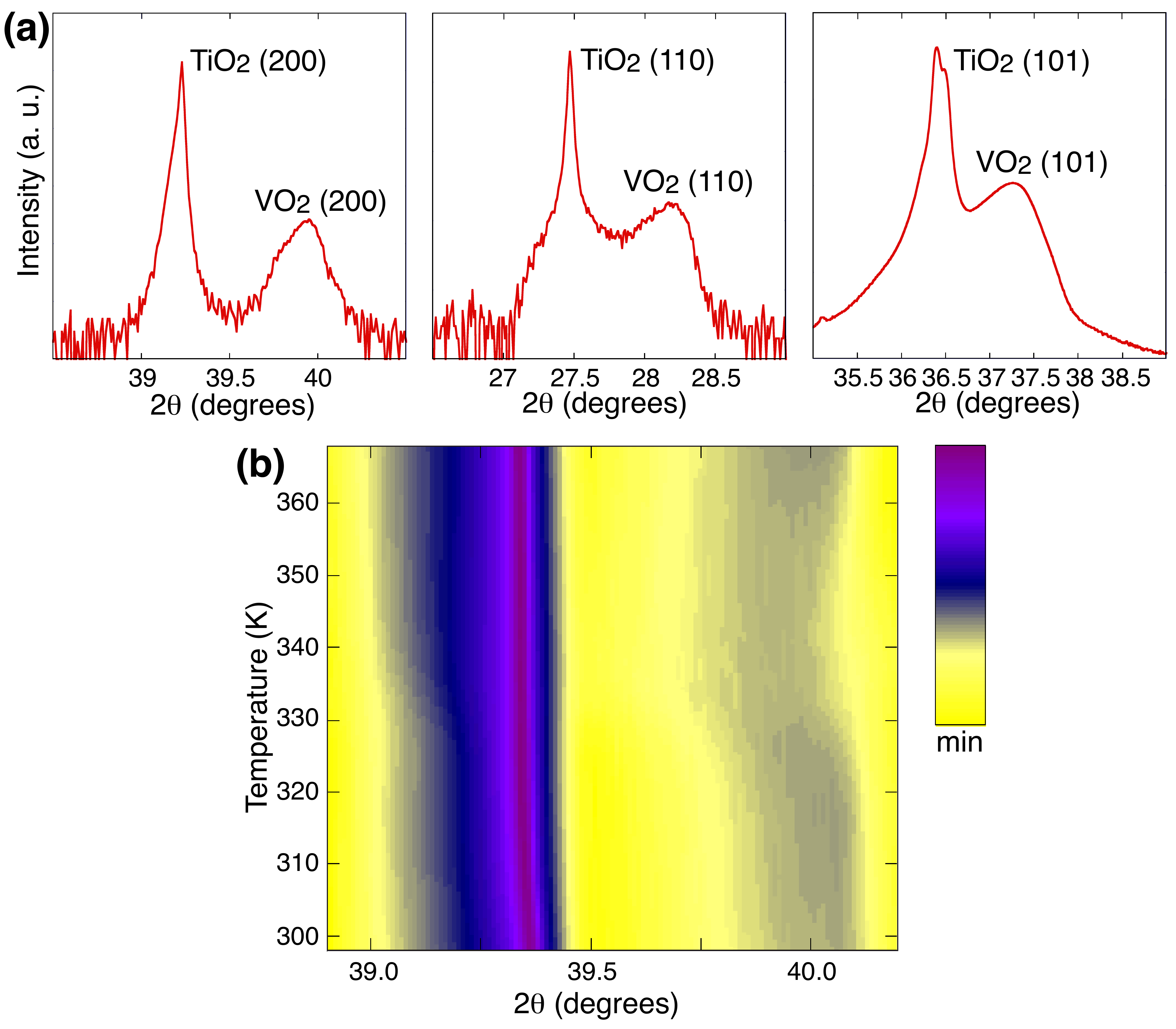}
\caption{(a) Room temperature XRD  of the $100nm$ thick VO$_2$ thin film, along three different directions: (200)$_R$, (110)$_R$, (101)$_R$. The lattice parameters can be estimated as $4.52\mathring{A}$ along a$_R$, $4.46\mathring{A}$ along b$_R$ and $2.89\mathring{A}$ along c$_R$, yielding mismatches of $-0.83\%$ along a$_R$, $-2.17\%$ along b$_R$ and $1.41\%$ along c$_R$. (b) Temperature dependent XRD of the $100nm$ thick VO$_2$ thin film. Both the TiO$_2$ and the VO$_2$ (200)$_R$ peaks are seen to shift as a function of temperature. There is no evidence of the development of any additional structural phase.}
\label{XRD}
\end{center}
\end{figure}

Detailed room temperature XRD of the VO$_2$ film confirms the single crystal nature of the sample and yields the lattice parameters of the material (Fig. \ref{XRD}(a)). Comparing these values with those in the rutile phase of bulk VO$_2$ \cite{McWhan1974} yields mismatches of $-0.83\%$ along a$_R$, $-2.17\%$ along b$_R$, and $1.41\%$ along c$_R$, indicative of a compressive strain along a$_R$ and b$_R$, as opposed to a tensile strain along c$_R$. Such strain values along c$_R$ are comparable to those achieved in one-dimensional VO$_2$ nanobeams \cite{Cao2009}.
Bulk rutile TiO$_2$ has larger lattice constants than bulk rutile VO$_2$ along all directions so one would expect that both b$_R$ and c$_R$ would expand in (100)$_R$ VO$_2$ films grown on a (100)$_R$ TiO$_2$ substrate. However, our nanosheets show that the expansion along c$_R$ surpasses the substrate clamping effect due to the b$_R$-axis, leading to a compressive strain in the ab$_R$-plane. This behavior is also observed in the $250nm$ thick film, with mismatches of $-0.68\%$ along a$_R^{250nm}$, $-1.94\%$ along b$_R^{250nm}$ and $0.86\%$ along c$_R^{250nm}$.

Temperature dependent XRD results (Fig. \ref{optCharacterization}(c)) show that a small a$_R$-axis expansion, $\sim0.1\%$, occurs during the film's structural transition. This transition occurs at a temperature $T_{St}^{film} \sim 340K \sim T_{St}^{bulk}$, the same as in bulk, and shows the expected hysteric behavior. (From here on $T_{St}$ will refer to both bulk and film structural transition temperatures.) The a$_R$-axis expansion in our strained sheet can be compared to that in bulk VO$_2$, estimated as $\frac{a_R-b_M}{b_M}\simeq 0.6-0.8\%$ \cite{Andersson1956, Kucharczyk1979}, where b$_M$ is the equivalent of a$_R$ in the bulk monoclinic insulating structure. The order of magnitude difference between the lattice parameter variation in the film and that in bulk stems from the clamping effect of the rutile TiO$_2$ substrate. This is consistent with the large strain observed in room temperature XRD, caused by the strong substrate clamping effect. The structural change exhibited at $340K$ by our strained VO$_2$ nanosheets is thus smaller than the change observed in bulk VO$_2$. Since VO$_2$ shares the TiO$_2$ rutile structure at high temperatures, the reduced structural change is expected to have an impact mainly on the low temperature VO$_2$ film's structure. The temperature dependent (200)$_R$ $2\theta$ plots from which the data in Fig. \ref{optCharacterization}(c) were extracted are presented in Fig. \ref{XRD}(b). These raw data provide further support to our observation that the clamping effect due to the substrate is very strong, thereby preventing the development of significantly different structural phases in the strained VO$_2$ film.

\begin{center}
\textbf{B. THz Time Domain Spectroscopy}
\end{center}

As shown in Fig. \ref{waveForm}, upon increasing the temperature from the insulating to the metallic phase the THz peak transmission in the $100nm$ (100)$_R$ VO$_2$ film drops by $\sim$70$\%$ along the  c$_R$-axis and by $\sim$15$\%$ along the b$_R$-axis, and the THz peak transmission in the $250nm$ (100)$_R$ VO$_2$ film drops by $\sim$85$\%$ along the  c$_R$-axis and by $\sim$15$\%$ along the b$_R$-axis.
Figure \ref{waveForm}(b) shows the transmission anisotropy in our $250nm$ VO$_2$ sample, normalized to its low temperature value along each axis (b$_R$ and c$_R$). This representation highlights the dramatic difference between the low temperature transmission along b$_R$ and that along c$_R$, thereby illustrating the potential of strained VO$_2$ films as temperature tunable THz polarizing beamsplitters.

\begin{figure} [ptb]
\begin{center}
\includegraphics[width=0.5\textwidth,keepaspectratio=true]{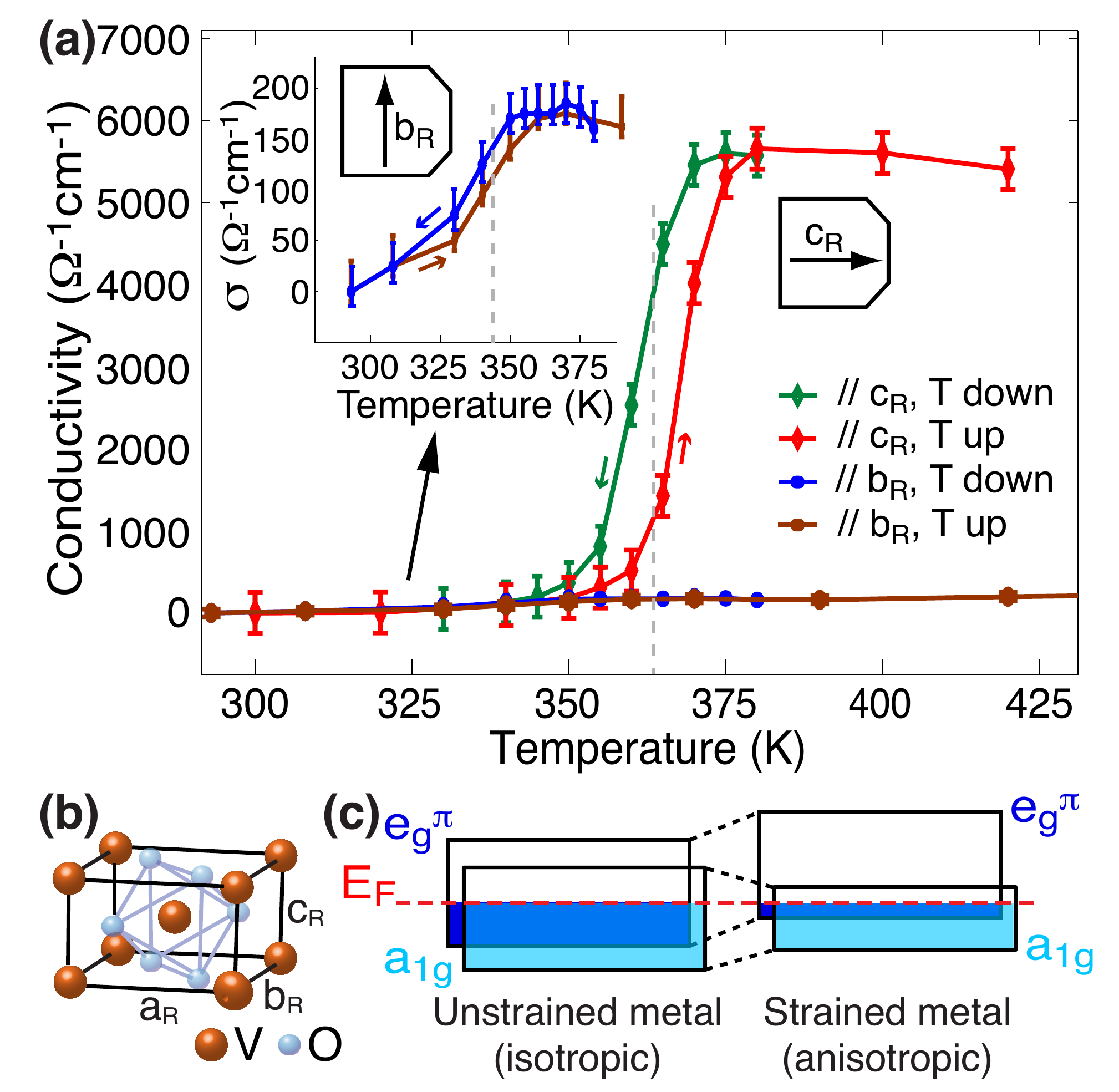}
\caption{(a) Temperature dependence of the far infrared conductivity in $100nm$ (100)$_R$ VO$_2$: $\sigma_{c_R}\simeq 30 \sigma_{b_R}$ above the MIT temperature; $T_{MI}^{c_R}=365K$ while $T_{MI}^{b_R}=340K = T_{MI}^{bulk}$. (b) The VO$_2$ rutile unit cell, following Eyert \cite{Eyert2002}. (c) Effect of tensile strain along c$_R$ on the electronic structure of metallic VO$_2$: antibonding e$_g^\pi$ bands are shifted upwards while the  a$_{1g}$ band is narrowed.}
\label{condData}
\end{center}
\end{figure}

Figure \ref{condData}(a) shows the temperature dependent far infrared conductivity, obtained from the transmission data, for the $100nm$ (100)$_R$ VO$_2$ nanosheet. The conductivity along c$_R$ shows a clear transition from the insulating to the metallic state with a narrow hysteresis, which is indicative of the high quality of the film. In the metallic state the conductivity is $\sigma_{c_R}\sim5650(\Omega cm)^{-1}$, comparable to bulk single crystal values \cite{Ladd1969}. The MIT along c$_R$ occurs at a temperature $T_{MI}^{c_R} \simeq 365^\circ K$. $T_{MI}^{c_R}$ is significantly larger than both the structural transition temperature and the bulk MIT temperature, $T_{St} \sim T_{MI}^{bulk} \sim 340K$. Our VO$_2$ films therefore exhibit, along c$_R$, a $\sim 25K$ temperature difference between the structural and the metal-insulator transition temperatures. The combination of the quasi three dimensionality of our nanosheets, which enables a direct measurement of the strain along the three crystal axes through XRD analysis, with the polarization sensitivity of THz spectroscopy is the key to identifying this distinction between the two transition temperatures.

The conductivity along b$_R$ also exhibits a transition (see inset of Fig. \ref{condData}(a)), which occurs at $T_{MI}^{b_R}\simeq 340^\circ K$. Along this direction we therefore observe that the structural and metal-insulator transition temperatures are the same, $T_{MI}^{b_R} \sim T_{St} \sim T_{MI}^{bulk}$. However, the conductivity along b$_R$ reaches a high temperature value about 30 times smaller than the high temperature conductivity along c$_R$. This strong conductivity anisotropy will be addressed later in the text.

\begin{figure} [ptb]
\begin{center}
\includegraphics[width=0.5\textwidth,keepaspectratio=true]{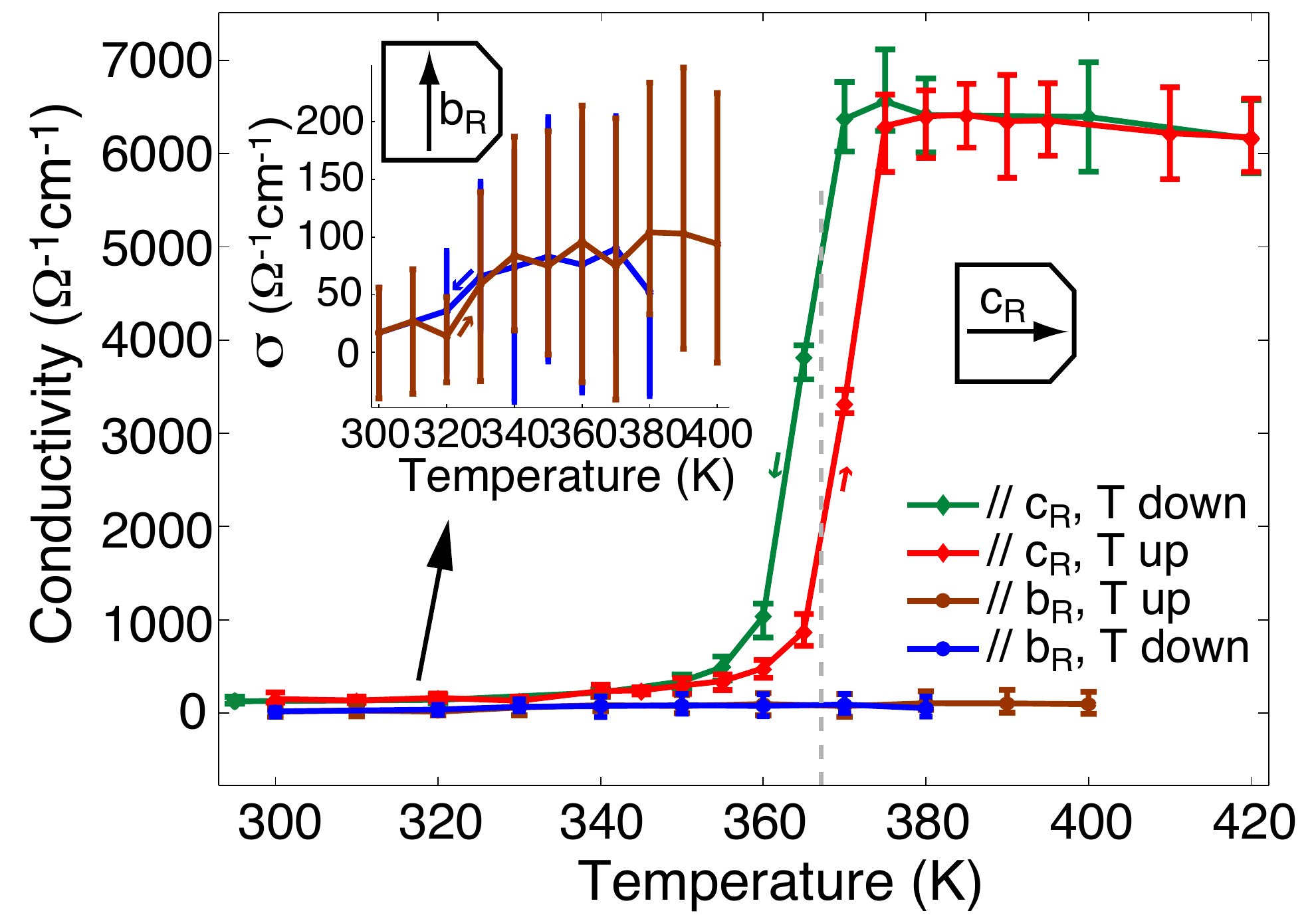}
\caption{Temperature dependence of the far-infrared conductivity in $250nm$ (100)$_R$ VO$_2$: $\sigma_{c_R}\simeq 5250(\Omega cm)^{-1}$ for $T>T_{MI}^{c_R}\simeq365K$; $\sigma_{b_R}<100(\Omega cm)^{-1}$ and $T_{MI}^{b_R}$ can only be estimated as $T_{MI}^{b_R}\simeq 340K$.}
\label{sample250VO2}
\end{center}
\end{figure}

THz TDS results for the $250nm$ thick sample (Fig. \ref{sample250VO2}) indicate that its transport properties  are similar to that of the $100nm$ sample. In particular, the high temperature conductivity along c$_R$ remains as good as in bulk VO$_2$ single crystals \cite{Ladd1969} while $T_{MI}^{c_R}\simeq365K>T_{MI}^{bulk}$. The high temperature conductivity along b$_R$ is very low, $<100(\Omega cm)^{-1}$, and the transition temperature can only be estimated at $T_{MI}^{b_R}\simeq340K$, consistent with $T_{MI}^{bulk}$ and $T_{St}$ and in line with what is observed in the $100nm$ sample (Fig. \ref{condData}(a)).

\begin{center}
\textbf{IV. DISCUSSION}
\end{center}

Understanding the contribution of the V3d orbitals to the electronic properties is crucial in order to explain the large material anisotropy in $T_{MI}$ \cite{Goodenough1971}. A splitting of the 5-fold degenerate 3d states occurs due to the octahedral coordination of the V atoms, resulting in a higher energy doubly degenerate $e_g$ level and a lower energy triply degenerate
$t_{2g}$ level. Trigonal distortion further splits the $t_{2g}$ levels leading to an upshifted doubly degenerate $e_g^\pi$ state,
responsible for conduction in the ab$_R$-plane, while downshifting a non-degenerate c$_R$-oriented $a_{1g}$ state (Fig. \ref{condData}(b)). Recent cluster Dynamical Mean Field Theory (cDMFT) calculations \cite{Lazarovits2010}, which include the effect of a $\pm 2\%$ strain along c$_R$, have demonstrated that a tensile strain along c$_R$ narrows the $a_{1g}$ derived band and leads to a compressive strain in the ab$_R$-plane, which uplifts the $e_g^\pi$ band (Fig. \ref{condData}(c)). In the Mott picture the energy increase of $e_g^\pi$ electrons reduces the screening of electrons residing in the $a_{1g}$ band, thereby enhancing the effect of correlations (i.e. increasing the screened Hubbard U). This results in an increase of the insulating band gap which opens, below T$_{MI}$, between the bonding $a_{1g}$ and the anti-bonding $e_g^\pi$ levels, therefore leading to an increased $T_{MI}$ along c$_R$. \cite{Eyert2002, Zylbersztejn1975}.

The epitaxial strain in our film can be decomposed into a uniaxial tensile strain along c$_R$ and a uniaxial compressive strain along b$_R$. Along c$_R$, $T_{MI}^{c_R}\sim 365K> T_{MI}^{bulk}$, in line with previous experimental results \cite{Muraoka2002}, while $\sigma_{c_R}$ remains comparable to the best single crystal values \cite{Ladd1969}. These results agree with the Mott picture above, where the increased lattice spacing along c$_R$ further increases the ratio of Coulomb repulsion to the inter-atomic hopping integral, thereby destabilizing the metallic phase and increasing $T_{MI}^{c_R}$. The Peierls picture predicts a decrease in $T_{MI}^{c_R}$ with tensile strain along c$_R$, thus failing to describe our results along that axis \cite{Lazarovits2010}. Also contrary to the Peierls-driven MIT scenario along c$_R$ is the fact that the structural transition occurs at a temperature $T_{St}\sim 340K$, $25K$ lower than $T_{MI}^{c_R}$: in a Peierls picture both transitions would be expected to occur at the same temperature. In contrast to what happens along c$_R$, the results along b$_R$, where $T_{MI}^{b_R} = T_{St}$, are compatible with a Peierls-driven MIT scenario.

As mentioned above, our samples are cracked along c$_R$. The occurrence of such cracks is common in VO$_2$, in both bulk and strained samples \cite{Berglund1969, Nagashima2006}, and this has prevented accurate measurements of the DC conductivity in this material in the direction perpendicular to c$_R$. Hindered quasiparticle motion along b$_R$ due to the presence of c$_R$ oriented cracks is the most likely explanation for the reduced value of $\sigma_{b_R}$ in our films. However, a different scenario could arise in the THz range. Due to the small value of the far-infrared carrier mean free path ($\sim \mathring{A}$ \cite{Qazilbash2006}) compared to DC,  the $\sim 1ps$ long THz field should be able to couple to the material along b$_R$ despite the cracks along c$_R$. According to this picture, the uniaxial compressive strain along b$_R$ would play a significant role in reducing $\sigma_{b_R}$, compared to $\sigma_{c_R}$, while keeping $T_{MI}^{b_R}=T_{MI}^{bulk}=T_{St}$. Given the orbitals orientation, conductivity in the ab$_R$-plane is mediated by the $e_g^\pi$ orbitals whose energy is controlled by the overlap between O2p and V3d orbitals \cite{Eyert2002}. Low conductivity behavior in the compressed ab$_R$-plane for $T>T_{MI}$ would then arise from the $e_g^\pi$ states being higher in energy than in the unstrained case, which reduces their overlap with the Fermi level (Fig. \ref{condData}C). VO$_2$ nanosheets that remain uncracked while maintaining a high level of strain along b$_R$ must be investigated in order to clarify the role of the $e_g^\pi$
orbitals on the high temperature value of $\sigma_{b_R}$, thereby also clarifying the nature of the MIT along that axis.

Our strained VO$_2$ films on a TiO$_2$ substrate have been shown to enable the separation of the far-infrared signal's polarization components both in time and intensity. The temperature dependent THz conductivity of strained VO$_2$ nanosheets (Figs. \ref{condData}(a) and \ref{sample250VO2}) shows that incident light polarized parallel to b$_R$ is transmitted through VO$_2$ at both low and high temperatures, whereas incident light polarized parallel to c$_R$ is transmitted through VO$_2$ at low temperatures but reflected at high temperatures. Our strained VO$_2$ single crystal nanosheets can thus be thought of as temperature switchable far-infrared polarizing beamsplitters.

\begin{center}
\textbf{V. PHASE DIAGRAM}
\end{center}

\begin{figure} [ptb]
\begin{center}
\includegraphics[width=0.5\textwidth,keepaspectratio=true]{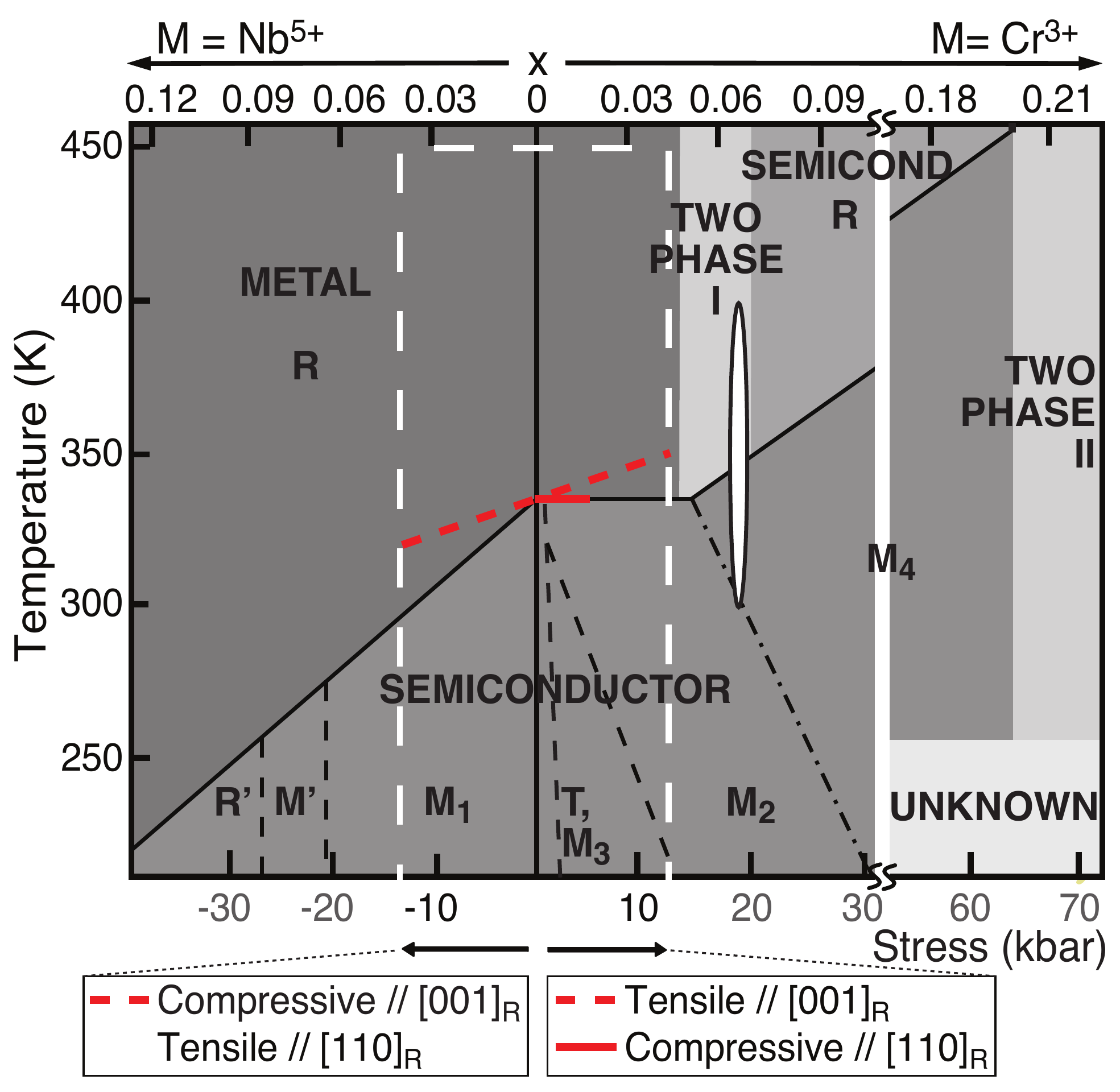}
\caption{Phase diagram of VO$_2$ for stress applied mostly along [110]. The black solid line indicates $T_{MI}$; the white oval defines the phase space of our samples; uniaxial stress results are presented in the region delimited by the dashed white lines; doping (V$_{1-x}$M$_x$O$_2$) leading to reduction (M=Nb$^{5+}$) or oxidation (M=Cr$^{3+}$) of V$^{4+}$ is specified on the top horizontal axis \cite{Pouget1975, Cao2009, Goodenough1973, Marezio1972, Pouget1976, Villeneuve1973} (refer to the main text for more details).}
\label{phaseDiagram}
\end{center}
\end{figure}

Finally, it is interesting to map the strain of our film onto the VO$_2$ phase diagram in Fig. \ref{phaseDiagram}, constructed based on past studies of VO$_2$ as a function of temperature, pressure \cite{Pouget1975, Cao2009} and doping \cite{Goodenough1973, Marezio1972, Pouget1976, Villeneuve1973}. In general, it is not trivial to map the effect of doping onto that of pressure. This was initiated by Pouget \textit{et al.}, who found a one-to-one correspondence between Cr-doping and the
application of uniaxial stress along [110]$_R$ \cite{Pouget1975, Pouget1976}. Other doping experiments tend to indicate that doping ions with radii smaller than V$^{4+}$ lead to an increase in $T_{MI}$ whereas those with larger radii have the opposite effect \cite{MacChesney1969}. This is similar to what happens in V$_2$O$_3$, a canonical Mott insulator \cite{McWhan1969}.

Figure \ref{phaseDiagram} compiles previous experimental results obtained on VO$_2$ under different pressure \cite{Pouget1975, Cao2009} and doping  \cite{Goodenough1973, Marezio1972, Pouget1976, Villeneuve1973} conditions.
$P=0$ corresponds to atmospheric pressure. The application of hydrostatic pressure \cite{Neuman1964, Berglund1969Hyd, Ladd1969} cannot be easily interpreted in terms of its influence on the conductivity nor on $T_{MI}$ along individual axes of the crystalline structure, which corresponds to the main focus of our study. We therefore do not include hydrostatic pressure data  in the phase diagram of Fig. \ref{phaseDiagram}. The dashed white lines delimit the region of the phase diagram where uniaxial pressure was applied along [001]$_R$ (up to $12kbar$), both for compression ($P<0$) and tension ($P>0$), and along [110]$_R$ (up to $\sim1.2kbar$), only for compression ($P>0$). Within the rectangle, the dashed red line represents $T_{MI}^{c_R}$ versus stress along c$_R$, [001]$_R$, while the solid red line represents $T_{MI}^{c_R}$ versus stress in the ab$_R$-plane, [110]$_R$. The dashed black lines separate different insulating phases within the low temperature monoclinic phase. The dash-doted line between phases M$_2$ and M$_4$ is a conjecture \cite{Goodenough1973, Villeneuve1973}. Note that apart from the dashed red line, relative to stress along [001]$_R$, all the information in the diagram refers to stress along [110]$_R$.
The white oval indicates the phase space characteristic of our (100)$_R$ VO$_2$ samples. The uniaxial stress on our films along [110]$_R$ has been estimated, respectively for the $100nm$ and the $250nm$ samples, to $\sim 20.9kbar$ and $\sim 18.5kbar$, based on the measured $-2.32\%$ and $-2.06\%$ mismatch along (110)$_R$, and on a Young modulus of $\sim 900kbar$ \cite{Sepulveda2008}. These uniaxial stress values are higher than ever reported, to our knowledge, for uniaxial stress on VO$_2$.

The dependence of $T_{MI}$ on doping, studied in V$_{1-x}$M$_x$O$_2$ compounds, can be approximately mapped onto its dependence on applied pressure. Reduction of V$^{4+}$ is achieved using M = Nb, Mo, W, Ta, Re, Ir, F, Ti, Os, Ru,  Tc, etc., with formal charges of +4, +5 or +6 \cite{Pouget1976, MacChesney1969, Goodenough1971, Beteille1998, Villeneuve1972, Piccirillo2007, Horlin1973, Holman2009, Nygren1969, Chae2008, Savborg1977, Bayard1975, Kristensen1968, Marinder1957, Gu2007}. The effect on $T_{MI}$ is similar to that of negative (compressive) stress along [001]$_R$, yielding $dT_{MI}/dx = -0.3 \sim -28K/at.\%M$. Oxidation of V$^{4+}$ is achieved using M = Cr, Al, Fe, Ga, Ge, Sn, Mn, Co, etc., with formal charges of +3 or +4 \cite{Pouget1976, MacChesney1969, Goodenough1971, Beteille1998, Marezio1972, Villeneuve1973, Drillon1974, Kosuge1967, Pollert1976, Pintchovski1978, Kitahiro1967, Lee1996}.  The effect on $T_{MI}$ is similar to that of positive (tensile) stress along [001]$_R$, yielding $dT_{MI}/dx = 0 \sim 13K/at.\%M$.

Our results, indicated by the white oval in the phase diagram of Fig. \ref{phaseDiagram}, are comparable to those of Everhart \textit{et al.}, where an anisotropy of about two orders of magnitude is observed in the metallic phase of bulk VO$_2$ single crystals doped with iron at $0.076\%$ \cite{Everhart1968}. Our estimate of the stress puts our samples in a region of the phase diagram where T$_{MI}$ increases with stress, and where a two phase behavior is expected above T$_{MI}$. This is consistent with our observation of both insulating ($T<T_{MI}^{c_R}$) and metallic ($T>T_{MI}^{c_R}$) values of $\sigma_{c_R}$ in the same rutile structure, above $T_{St}$. The independence of the conductivity behavior from the structural phase has been reported before \cite{Kim2006, Qazilbash2011, Arcangeletti2007}, though relative to monoclinic structures. As for the low temperature structure, M4 or rutile seem to be the most likely candidates, although we cannot distinguish between the two. M1, M2 and M3/T are all structurally too far from rutile \cite{Marezio1972, Goodenough1973, Jones2010} to lead to the results shown in Fig. \ref{optCharacterization}C, and they are located in different regions of the phase diagram, as seen in Fig.  \ref{phaseDiagram}. These observations, along with the strong conductivity anisotropy exhibited by our strained nanosheets, are indicative of a more complex behavior of VO$_2$, beyond the currently accepted doping and strain dependence of its structural and transport properties (Fig. \ref{phaseDiagram}).

Coming back to the conductivity measured in our strained nanosheets, it is important to point out that its anisotropy is unexpectedly large compared to previous experimental observations and theoretical calculations on VO$_2$ samples \cite{Lu2008, Barker1966, Everhart1968, Bongers1965, Kosuge1967}. The anisotropy in the DC conductivity, $\frac{\sigma_{c_R}}{\sigma_{b_R}}$ or $\frac{\sigma_{c_R}}{\sigma_{a_R}}$, in undoped VO$_2$ samples is
generally $>$1 for $T<T_{MI}$ but can take many different values for $T>T_{MI}$ (Table \ref{tableS1}). Differences in sample quality and stoichiometry as well as in conductivity measurement techniques surely affect the results, but variations in geometry and internal strain / cracking are likely to also have an effect on the anisotropy of metallic VO$_2$.

Tables \ref{tableS1} and \ref{tableS2} present a complete and up to date review, to our knowledge, of the experimental and theoretical data on conductivity anisotropy in VO$_2$, including the effect of externally applied stress. Table \ref{tableS1} includes the results for the conductivity anisotropy, above and below T$_{MI}$, when no external stress is applied \cite{Bongers1965, Barker1966, Kosuge1967, Everhart1968, Verleur1968, Continenza1999, Mossanek2007, Lysenko2007, Lu2008, Tomczak2009}. Table \ref{tableS2} presents several results for the strain induced variation of T$_{MI}$ and of the conductivity in VO$_2$, for situations where hydrostatic or uniaxial pressure is applied to the samples \cite{Minomura1964, Neuman1964, Berglund1969Hyd, Ladd1969, Pouget1975, Gregg1997, Muraoka2002, Arcangeletti2007, Lu2008, Sakai2008, Cao2009}.

As seen in tables \ref{tableS1} and \ref{tableS2}, in most situations where uniaxial pressure is applied the conductivity is measured along c$_R$, the only axis along which it is well defined due to geometry constraints of the samples or to cracking. Few studies determine the conductivity along a$_R$ or b$_R$ as a function of applied pressure. In general, it is agreed that
(i) an applied uniaxial compressive (tensile) stress along the c$_R$-axis leads to an increased (decreased) conductivity at $T>T_{MI}$;
(ii) an applied uniaxial compressive (tensile) stress along the c$_R$-axis leads to a decreased (increased) $T_{MI}$ (the axis along which $T_{MI}$ is measured is not always specified);
(iii) a small uniaxial stress applied along the [110]$_R$ direction has no significant effect on the conductivity nor $T_{MI}$ along c$_R$ but it promotes a phase transition between different monoclinic structures (M1, M2, T/M3, M4), in the insulating phase \cite{Pouget1975, Marezio1972, Goodenough1973}.
Further experiments are needed in order to systematically measure the conductivity along a$_R$ or b$_R$ under
(i) an applied uniaxial stress along a$_R$ or b$_R$;
(ii) an applied uniaxial stress along c$_R$;
(iii) hydrostatic pressure.
Also, clear criteria for distinguishing $T_{MI}$ from $T_{St}$ would be extremely valuable to help draw a more accurate and complete phase diagram, in line with what has been attempted in previous studies \cite{Kim2006, Qazilbash2011, Arcangeletti2007}. The highly oriented strain across a quasi three dimensional structure, achieved in epitaxially grown VO$_2$ nanosheets, offers an extraordinary versatility and potential for investigation of these issues.

Theoretically, several pictures have been suggested to explain the anisotropy in (unstrained) VO$_2$:
(1) a two band model description of 3d electrons, within the framework of the Goodenough model, predicts a non conducting ab$_R$-plane for $T>T_{MI}$, any residual conductivity in that plane being due to the overlap of O2p and V3d orbitals \cite{Goodenough1971, Eyert2002};
(2) LDA calculations by Allen's group predict the structural distortion to be the main force driving the MIT in VO$_2$, following a simple Peierls picture \cite{Wentzcovitch1994};
(3) a three-band Hubbard model, suggested by Tanaka \textit{et al.}, predicts a one dimensional conducting phase along c$_R$ for $T<T_{MI}$ and an isotropically conducting phase for $T>T_{MI}$ \cite{Tanaka2003};
(4) LDA+cDMFT calculations by Biermann \textit{et al.} suggest that electron correlations within the e$_g^{\pi}$ levels are weaker than those along the a$_{1g}$ ones \cite{Biermann2005};
(5) subsequent LDA+cDMFT calculations by Kotliar's group, which include a moderate degree of uniaxial strain, suggest that electronic correlations are the main driving force in the MIT, and that the rutile phase itself should be able to
support both metallic and insulating electronic behavior \cite{Lazarovits2010};
(6) Liebsch \textit{et al.} compared the LDA+U, DMFT and GW methods and found that none of them provides a full description of VO$_2$, namely of the development of the insulating gap below T$_{MI}$ \cite{Liebsch2005}.
Nonetheless, most of the calculations based on unstrained VO$_2$ samples agree that
(i) a$_{1g}$ and e$_g^\pi$ states can be regarded as nearly independent;
(ii) the conductivity is expected to be of predominantly a$_{1g}$ behavior for $T<T_{MI}$ and nearly isotropic for $T>T_{MI}$;
(iii) electron correlations should be included in the model, mainly in the c$_R$ oriented a$_{1g}$ levels.
Our results agree with (i) and (iii), while strain induced cracking prevents the accurate description of the effect mentioned in (ii).
Further theoretical investigations should take into account not only strain \cite{Lazarovits2010}  but also the oxygen degrees of freedom, in order to provide a more accurate description of the O2p and V3d orbitals overlap, most relevant for describing the conductivity in the ab$_R$-plane, and of the subsequent cracking along c$_R$. Such estimates would be adequate for a quantitative, rather than merely qualitative, comparison with our experimental data.

\begin{center}
\textbf{VI. CONCLUSION}
\end{center}

In summary, we have observed a large anisotropy of the properties of strained $100nm$ and $250nm$ thick VO$_2$ nanosheets. The increased value of $T_{MI}^{c_R}$ compared to $T_{St}$ is a clear signature of a Mott- rather than Peierls-driven MIT along c$_R$. An $e_g^\pi$ orbital tuning picture is proposed to explain the reduced value of the high temperature $\sigma_{b_R}$, although a crack induced conductivity decrease cannot be conclusively ruled out. Additional experiments are needed in order to clarify the mechanism of the phase transition in strained VO$_2$, mainly in the direction perpendicular to c$_R$. This would allow the development of a more comprehensive phase diagram for this material. In general, epitaxial strain engineering is a powerful tool which has the potential to enable careful tuning of the metal-insulator transition in numerous other correlated electron materials, thereby providing a viable route towards technologically relevant multifunctionality and increased understanding of the microscopic origin of the MIT.

\begin{center}
\textbf{ACKNOWLEDGMENTS}
\end{center}

We thank Kebin Fan and Wei Zhang for SEM sample characterization.
We acknowledge support from DOE - Basic Energy Sciences under Grant No. DE-FG02-09ER46643 (E.A., M.L., R.D.A.). E. Abreu acknowledges support from Funda\c c\~ao para a Ci\^encia e a Tecnologia, Portugal, through a doctoral degree fellowship (SFRH/ BD/ 47847/ 2008).

\newpage

\begin{table}
\renewcommand{\arraystretch}{0.7}
\begin{tabular}{| c | c | c | c | c |}
\hline
\textbf{publication} & \textbf{sample type} & \textbf{method} & \textbf{anisotropy} & \textbf{anisotropy}\\
 & & & \textbf{below $T_{MI}$} & \textbf{above $T_{MI}$}\\
\hline
\hline
Bongers \textit{et al.} & bulk single crystal & two-probe & 2 & 2.5\\
(1965) & (needle along c$_R$, &  &  &  \\
 & 6x0.3x0.15$mm$) &  &  &  \\
\hline
Barker \textit{et al.} & bulk polycrystal & Hall voltage & 0.7-0.8 & 1.1-2\\
(1966) & (well defined c$_R$) & & &\\
\hline
Kosuge \textit{et al.} & bulk single crystal & two-probe & 1.14 & 0.58\\
(1967) & & & & \\
\hline
Koide \textit{et al.} & single crystal film & two probe & 2 & 0.001-0.1\\
(1967) & on rutile substrate & & & \\
 \cline{2-5}
   & bulk single crystal & two-probe & $<1.2$ & 0.33\\
   & (needle along c$_R$, &  &  &  \\
   & 3x0.8x0.07$mm$) & & & \\
\hline
Everhart \textit{et al.} & bulk single crystal & four-probe & 2-10 & 7.5\\
(1968) & (4-7x1-4x1-4$mm$) & & & \\
\hline
Verleur \textit{et al.} & bulk single crystal & reflectivity & 0.28-1.4 & 0.79-2\\
(1968) & & (0.25-$5eV$) & & \\
\hline
Continenza \textit{et al.} & (calculations) & model GW & 3-6.7 & N/A\\
(1999) & & (0-10$eV$) & & \\
\hline
Mossanek \textit{et al.} & (calculations) & LDA (0-12$eV$) & 0.6-$>3$ & 0.7-3.3\\
(2007) & & & & \\
\hline
Lysenko \textit{et al.} & 30$nm$ film on & optical (400$nm$) & 1.05 & 1\\
(2007) & (012) Al$_2$O$_3$ & diffraction & & \\
\hline
Lu \textit{et al.} & 40$nm$ film on & star-shaped & 5.14 & 1\\
(2008) & (011)$_R$ TiO$_2$ & electrodes & & \\
\hline
Tomczak et al. & (calculations) & LDA+CDMFT & 0.73-1.67 & 0.96-1.14\\
(2009) & & (0-5$eV$) & & \\
\hline
our results & 100$nm$ film on & THz TDS & N/A & $\sim$30\\
(2010) & (100)$_R$ TiO$_2$ & (far infrared) & & \\
\hline
\end{tabular}
\caption{Compilation of previous experimental and theoretical results for the conductivity (DC, unless otherwise specified) anisotropy, $\frac{\sigma_{c_R}}{\sigma_{b_R}}$ or $\frac{\sigma_{c_R}}{\sigma_{a_R}}$, in VO$_2$, above and below T$_{MI}$, when no external stress is applied. \cite{Bongers1965, Barker1966, Kosuge1967, Everhart1968, Verleur1968, Continenza1999, Mossanek2007, Lysenko2007, Lu2008, Tomczak2009}}
\label{tableS1}
\end{table}

\begin{table}
\renewcommand{\arraystretch}{0.7}
\scalebox{0.8}{%
\begin{tabular}{| c | c | c | c | c | c | c |}
\hline
\textbf{publication} & \textbf{sample} & \textbf{applied} & \textbf{conductivity} & \textbf{conductivity} & \textbf{$dT_{MI}/dP$} & \textbf{$dT_{MI}/dP$}\\
 & \textbf{type} & \textbf{pressure} & \textbf{vs. pressure} & \textbf{along...} & \textbf{[K/kbar]} & \textbf{along...} \\
\hline
\hline
Minomura \textit{et al.} & bulk & uniaxial & not specified & not specified & -0.46 & not specified\\
(1964) & & (Drickamer cell, & & & & \\
 & & up to 160kbar) & & & & \\
\hline
Neuman \textit{et al.} & single crystal & hydrostatic & increases & not specified & no significant & not specified\\
(1964) & & (up to 6kbar) & below $T_{MI}$ & & change & \\
\cline{2-7}
 & powder & hydrostatic & increases & not specified & no significant & not specified\\
 & & (up to 6kbar) & & & change & \\
\hline
Berglund \textit{et al.} & bulk single crystal & hydrostatic & increases; & c$_R$  & 0.082 & c$_R$ \\
(1969) & (0.25x0.25x0.2in, & (up to 44kbar) & saturates above & & & \\
 & crack along c$_R$) & & $T_{MI}$, at 15-20kbar & & & \\
\hline
Ladd \textit{et al.} & bulk single crystal & hydrostatic & increases & c$_R$  & 0.06 &  c$_R$  \\
(1969) & (10x1x1mm) & (up to 8kbar) & (up to 30kbar, & & & \\
 & & & at T$_{room}$) & & & \\
\cline{3-7}
 & & along a$_R$ or b$_R$ & N/A & c$_R$ & no significant & c$_R$ \\
 & & & & & change & \\
\cline{3-7}
 & & along c$_R$ & increases & c$_R$ & -1.2 & c$_R$ \\
 & & (up to 0.5kbar) & below $T_{MI}$ & & & \\
\hline
Pouget \textit{et al.} & bulk single crystal & along [110]$_R$ & no significant & c$_R$ & no significant & c$_R$ \\
(1975) & (0.5mm$^2$x2mm, & (up to 1.2kbar) & change & & change & \\
 & 3mm$^2$x4mm) & & & & & \\
\hline
Gregg \textit{et al.} & thin films on & in-plane & increases & not specified & N/A & N/A \\
(1997) & Al$_2$O$_3$(012) &  (3-point bend) & below $T_{IM}$ & & & \\
\hline
Muraoka \textit{et al.} & 10-15nm thick & along c$_R$ & increases & a$_R$ or b$_R$ & $<0$ & a$_R$ or b$_R$ \\
(2002) & single crystal thin & (epitaxial: & & for [001]$_R$; & & for [001]$_R$; \\
 & films on (001)$_R$ & -0.3$\%$ for (001)$_R$, & & not specified & &  not specified \\
 & and (110)$_R$ TiO$_2$ & 1.2$\%$ for (110)$_R$) & & for [110]$_R$ & & for [110]$_R$ \\
\hline
Arcangeletti \textit{et al.} & single crystal & uniaxial & increases below & not specified & N/A & N/A \\
(2007) & (5$\mu$m thick slab) & (diamond a. c., & $T_{MI}$ & & & \\
 & & up to 140kbar) & (750-6000cm$^{-1}$) & & & \\
\hline
Lu \textit{et al.} & single crystal & in-plane & N/A & a$_R$ or b$_R$ & $<0$ &  a$_R$ or b$_R$ \\
(2008) & thin film on & (epitaxial: & & & & \\
 & (011)$_R$ TiO$_2$ &  -1.2$\%$//[011]$_R$, & & & & \\
 & & -0.4$\%$//[001]$_R$) & & & & \\
\hline
Sakai & polycrystalline thin & in-plane & increases & not specified & $>0$ & not specified \\
(2008) & film on metallic Ti & (point contact, & & & & \\
 & (a$_R$ in plane) & up to 255kbar & & & & \\
\hline
Cao \textit{et al.} & single crystal  & along c$_R$ & increases & c$_R$ & $\sim$-2 & c$_R$ \\
(2009) & (0.5-2x0.5-2x100 & (3-point bend, & (optical) & & & \\
 & nm, along c$_R$) & up to 12kbar) & & & & \\
\hline
our results & single crystal  & in-plane & no significant & c$_R$ & $<0$ & c$_R$ \\
(2010) & thin film on & (epitaxial: & change (far IR))& & & \\
\cline{4-7}
 & (100)$_R$ TiO$_2$ & $5.4\%$//[001]$_R$, & decreases above & b$_R$ & no significant & b$_R$ \\
 & & $-2.3\%$//[010]$_R$) & $T_{MI}$ (far IR) & & change & \\
\hline
\end{tabular}}
\caption{Compilation of previous experimental and theoretical results for the variation of $T_{MI}$ and of the conductivity (DC, unless otherwise specified) in VO$_2$, under applied hydrostatic or uniaxial pressures. We assume $\Delta P>0$ for compression and $\Delta P<0$ for expansion. \cite{Minomura1964, Neuman1964, Berglund1969Hyd, Ladd1969, Pouget1975, Gregg1997, Muraoka2002, Arcangeletti2007, Lu2008, Sakai2008, Cao2009}}
\label{tableS2}
\end{table}


\begin{thebibliography}{10}%
\makeatletter
\providecommand \@ifxundefined [1]{%
 \ifx #1\undefined \expandafter \@firstoftwo
 \else \expandafter \@secondoftwo
\fi
}%
\providecommand \@ifnum [1]{%
 \ifnum #1\expandafter \@firstoftwo
 \else \expandafter \@secondoftwo
\fi
}%
\providecommand \enquote [1]{``#1''}%
\providecommand \bibnamefont  [1]{#1}%
\providecommand \bibfnamefont [1]{#1}%
\providecommand \citenamefont [1]{#1}%
\providecommand\href[0]{\@sanitize\@href}%
\providecommand\@href[1]{\endgroup\@@startlink{#1}\endgroup\@@href}%
\providecommand\@@href[1]{#1\@@endlink}%
\providecommand \@sanitize [0]{\begingroup\catcode`\&12\catcode`\#12\relax}%
\@ifxundefined \pdfoutput {\@firstoftwo}{%
 \@ifnum{\z@=\pdfoutput}{\@firstoftwo}{\@secondoftwo}%
}{%
 \providecommand\@@startlink[1]{\leavevmode}%
 \providecommand\@@endlink[0]{}%
}{%
 \providecommand\@@startlink[1]{%
  \leavevmode
  \pdfstartlink
   attr{/Border[0 0 1 ]/H/I/C[0 1 1]}%
   user{/Subtype/Link/A<</Type/Action/S/URI/URI(#1)>>}%
  \relax
 }%
 \providecommand\@@endlink[0]{\pdfendlink}%
}%
\providecommand \url  [0]{\begingroup\@sanitize \@url }%
\providecommand \@url [1]{\endgroup\@href {#1}{\urlprefix}}%
\providecommand \urlprefix [0]{URL }%
\providecommand \Eprint[0]{\href }%
\@ifxundefined \urlstyle {%
  \providecommand \doi [1]{doi:\discretionary{}{}{}#1}%
}{%
  \providecommand \doi [0]{doi:\discretionary{}{}{}\begingroup
  \urlstyle{rm}\Url }%
}%
\providecommand \doibase [0]{http://dx.doi.org/}%
\providecommand \Doi[1]{\href{\doibase#1}}%
\providecommand \bibAnnote [3]{%
  \BibitemShut{#1}%
  \begin{quotation}\noindent
    \textsc{Key:}\ #2\\\textsc{Annotation:}\ #3%
  \end{quotation}%
}%
\providecommand \bibAnnoteFile [2]{%
  \IfFileExists{#2}{\bibAnnote {#1} {#2} {\input{#2}}}{}%
}%
\providecommand \typeout [0]{\immediate \write \m@ne }%
\providecommand \selectlanguage [0]{\@gobble}%
\providecommand \bibinfo [0]{\@secondoftwo}%
\providecommand \bibfield [0]{\@secondoftwo}%
\providecommand \translation [1]{[#1]}%
\providecommand \BibitemOpen[0]{}%
\providecommand \bibitemStop [0]{}%
\providecommand \bibitemNoStop [0]{.\EOS\space}%
\providecommand \EOS [0]{\spacefactor3000\relax}%
\providecommand \BibitemShut [1]{\csname bibitem#1\endcsname}%
\bibitem{Ladd1969}%
  \BibitemOpen
  \bibfield{author}{%
  \bibinfo {author} {\bibfnamefont{L.~A.}\ \bibnamefont{Ladd}}\ and\ \bibinfo
  {author} {\bibfnamefont{W.}~\bibnamefont{Paul}},\ }%
  \bibfield{journal}{%
  \bibinfo {journal} {Solid State Communications}\ }%
  \textbf{\bibinfo {volume} {7}},\ \bibinfo {pages} {425} (\bibinfo {year}
  {1969})%
  \bibAnnoteFile{NoStop}{Ladd1969}%
\bibitem{Andersson1956}%
  \BibitemOpen
  \bibfield{author}{%
  \bibinfo {author} {\bibfnamefont{G.}~\bibnamefont{Andersson}},\ }%
  \bibfield{journal}{%
  \bibinfo {journal} {Acta Chemica Scandinavica}\ }%
  \textbf{\bibinfo {volume} {10}},\ \bibinfo {pages} {623} (\bibinfo {year}
  {1956})%
  \bibAnnoteFile{NoStop}{Andersson1956}%
\bibitem{Biermann2005}%
  \BibitemOpen
  \bibfield{author}{%
  \bibinfo {author} {\bibfnamefont{S.}~\bibnamefont{Biermann}}, \bibinfo
  {author} {\bibfnamefont{A.}~\bibnamefont{Poteryaev}}, \bibinfo {author}
  {\bibfnamefont{A.}~\bibnamefont{Lichtenstein}},\ and\ \bibinfo {author}
  {\bibfnamefont{A.}~\bibnamefont{Georges}},\ }%
  \bibfield{journal}{%
  \Doi{10.1103/PhysRevLett.94.026404}{\bibinfo {journal} {Physical Review
  Letters}}\ }%
  \textbf{\bibinfo {volume} {94}},\ \bibinfo {pages} {026404} (\bibinfo {month}
  {Jan.}\ \bibinfo {year} {2005}),\ ISSN \bibinfo {issn} {0031-9007},\
  \url{http://link.aps.org/doi/10.1103/PhysRevLett.94.026404}%
  \bibAnnoteFile{NoStop}{Biermann2005}%
\bibitem{Lazarovits2010}%
  \BibitemOpen
  \bibfield{author}{%
  \bibinfo {author} {\bibfnamefont{B.}~\bibnamefont{Lazarovits}}, \bibinfo
  {author} {\bibfnamefont{K.}~\bibnamefont{Kim}}, \bibinfo {author}
  {\bibfnamefont{K.}~\bibnamefont{Haule}},\ and\ \bibinfo {author}
  {\bibfnamefont{G.}~\bibnamefont{Kotliar}},\ }%
  \bibfield{journal}{%
  \Doi{10.1103/PhysRevB.81.115117}{\bibinfo {journal} {Physical Review B}}\ }%
  \textbf{\bibinfo {volume} {81}},\ \bibinfo {pages} {115117} (\bibinfo {month}
  {Mar.}\ \bibinfo {year} {2010}),\ ISSN \bibinfo {issn} {1098-0121},\
  \url{http://link.aps.org/doi/10.1103/PhysRevB.81.115117}%
  \bibAnnoteFile{NoStop}{Lazarovits2010}%
\bibitem{Kim2006}%
  \BibitemOpen
  \bibfield{author}{%
  \bibinfo {author} {\bibfnamefont{H.-T.}\ \bibnamefont{Kim}}, \bibinfo
  {author} {\bibfnamefont{Y.~W.}\ \bibnamefont{Lee}}, \bibinfo {author}
  {\bibfnamefont{B.-J.}\ \bibnamefont{Kim}}, \bibinfo {author}
  {\bibfnamefont{B.-G.}\ \bibnamefont{Chae}}, \bibinfo {author}
  {\bibfnamefont{S.~J.}\ \bibnamefont{Yun}}, \bibinfo {author}
  {\bibfnamefont{K.-Y.}\ \bibnamefont{Kang}}, \bibinfo {author}
  {\bibfnamefont{K.-J.}\ \bibnamefont{Han}}, \bibinfo {author}
  {\bibfnamefont{K.-J.}\ \bibnamefont{Yee}},\ and\ \bibinfo {author}
  {\bibfnamefont{Y.-S.}\ \bibnamefont{Lim}},\ }%
  \bibfield{journal}{%
  \Doi{10.1103/PhysRevLett.97.266401}{\bibinfo {journal} {Physical Review
  Letters}}\ }%
  \textbf{\bibinfo {volume} {97}},\ \bibinfo {pages} {266401} (\bibinfo {month}
  {Dec.}\ \bibinfo {year} {2006}),\ ISSN \bibinfo {issn} {0031-9007},\
  \url{http://link.aps.org/doi/10.1103/PhysRevLett.97.266401}%
  \bibAnnoteFile{NoStop}{Kim2006}%
\bibitem{Qazilbash2011}%
  \BibitemOpen
  \bibfield{author}{%
  \bibinfo {author} {\bibfnamefont{M.}~\bibnamefont{Qazilbash}}, \bibinfo
  {author} {\bibfnamefont{A.}~\bibnamefont{Tripathi}}, \bibinfo {author}
  {\bibfnamefont{A.}~\bibnamefont{Schafgans}}, \bibinfo {author}
  {\bibfnamefont{B.-J.}\ \bibnamefont{Kim}}, \bibinfo {author}
  {\bibfnamefont{H.-T.}\ \bibnamefont{Kim}}, \bibinfo {author}
  {\bibfnamefont{Z.}~\bibnamefont{Cai}}, \bibinfo {author}
  {\bibfnamefont{M.}~\bibnamefont{Holt}}, \bibinfo {author}
  {\bibfnamefont{J.}~\bibnamefont{Maser}}, \bibinfo {author}
  {\bibfnamefont{F.}~\bibnamefont{Keilmann}}, \bibinfo {author}
  {\bibfnamefont{O.}~\bibnamefont{Shpyrko}},\ and\ \bibinfo {author}
  {\bibfnamefont{D.}~\bibnamefont{Basov}},\ }%
  \bibfield{journal}{%
  \Doi{10.1103/PhysRevB.83.165108}{\bibinfo {journal} {Physical Review B}}\ }%
  \textbf{\bibinfo {volume} {83}},\ \bibinfo {pages} {165108} (\bibinfo {month}
  {Apr.}\ \bibinfo {year} {2011}),\ ISSN \bibinfo {issn} {1098-0121},\
  \url{http://link.aps.org/doi/10.1103/PhysRevB.83.165108}%
  \bibAnnoteFile{NoStop}{Qazilbash2011}%
\bibitem{Arcangeletti2007}%
  \BibitemOpen
  \bibfield{author}{%
  \bibinfo {author} {\bibfnamefont{E.}~\bibnamefont{Arcangeletti}}, \bibinfo
  {author} {\bibfnamefont{L.}~\bibnamefont{Baldassarre}}, \bibinfo {author}
  {\bibfnamefont{D.}~\bibnamefont{{Di Castro}}}, \bibinfo {author}
  {\bibfnamefont{S.}~\bibnamefont{Lupi}}, \bibinfo {author}
  {\bibfnamefont{L.}~\bibnamefont{Malavasi}}, \bibinfo {author}
  {\bibfnamefont{C.}~\bibnamefont{Marini}}, \bibinfo {author}
  {\bibfnamefont{A.}~\bibnamefont{Perucchi}},\ and\ \bibinfo {author}
  {\bibfnamefont{P.}~\bibnamefont{Postorino}},\ }%
  \bibfield{journal}{%
  \Doi{10.1103/PhysRevLett.98.196406}{\bibinfo {journal} {Physical Review
  Letters}}\ }%
  \textbf{\bibinfo {volume} {98}},\ \bibinfo {pages} {196406} (\bibinfo {month}
  {May}\ \bibinfo {year} {2007}),\ ISSN \bibinfo {issn} {0031-9007},\
  \url{http://link.aps.org/doi/10.1103/PhysRevLett.98.196406}%
  \bibAnnoteFile{NoStop}{Arcangeletti2007}%
\bibitem{Kim2005}%
  \BibitemOpen
  \bibfield{author}{%
  \bibinfo {author} {\bibfnamefont{H.-T.}\ \bibnamefont{Kim}}, \bibinfo
  {author} {\bibfnamefont{B.-G.}\ \bibnamefont{Chae}}, \bibinfo {author}
  {\bibfnamefont{D.-H.}\ \bibnamefont{Youn}}, \bibinfo {author}
  {\bibfnamefont{G.}~\bibnamefont{Kim}}, \bibinfo {author}
  {\bibfnamefont{K.-Y.}\ \bibnamefont{Kang}}, \bibinfo {author}
  {\bibfnamefont{S.-J.}\ \bibnamefont{Lee}}, \bibinfo {author}
  {\bibfnamefont{K.}~\bibnamefont{Kim}},\ and\ \bibinfo {author}
  {\bibfnamefont{Y.-S.}\ \bibnamefont{Lim}},\ }%
  \bibfield{journal}{%
  \Doi{10.1063/1.1941478}{\bibinfo {journal} {Applied Physics Letters}}\ }%
  \textbf{\bibinfo {volume} {86}},\ \bibinfo {pages} {242101} (\bibinfo {year}
  {2005}),\ ISSN \bibinfo {issn} {00036951},\
  \url{http://link.aip.org/link/APPLAB/v86/i24/p242101/s1\&Agg=doi}%
  \bibAnnoteFile{NoStop}{Kim2005}%
\bibitem{Driscoll2009}%
  \BibitemOpen
  \bibfield{author}{%
  \bibinfo {author} {\bibfnamefont{T.}~\bibnamefont{Driscoll}}, \bibinfo
  {author} {\bibfnamefont{H.-T.}\ \bibnamefont{Kim}}, \bibinfo {author}
  {\bibfnamefont{B.-G.}\ \bibnamefont{Chae}}, \bibinfo {author}
  {\bibfnamefont{B.-J.}\ \bibnamefont{Kim}}, \bibinfo {author}
  {\bibfnamefont{Y.-W.}\ \bibnamefont{Lee}}, \bibinfo {author}
  {\bibfnamefont{N.~M.}\ \bibnamefont{Jokerst}}, \bibinfo {author}
  {\bibfnamefont{S.}~\bibnamefont{Palit}}, \bibinfo {author}
  {\bibfnamefont{D.~R.}\ \bibnamefont{Smith}}, \bibinfo {author}
  {\bibfnamefont{M.}~\bibnamefont{{Di Ventra}}},\ and\ \bibinfo {author}
  {\bibfnamefont{D.~N.}\ \bibnamefont{Basov}},\ }%
  \bibfield{journal}{%
  \Doi{10.1126/science.1176580}{\bibinfo {journal} {Science}}\ }%
  \textbf{\bibinfo {volume} {325}},\ \bibinfo {pages} {1518} (\bibinfo {month}
  {Sep.}\ \bibinfo {year} {2009}),\ ISSN \bibinfo {issn} {1095-9203},\
  \url{http://www.ncbi.nlm.nih.gov/pubmed/19696311}%
  \bibAnnoteFile{NoStop}{Driscoll2009}%
\bibitem{Hilton2007}%
  \BibitemOpen
  \bibfield{author}{%
  \bibinfo {author} {\bibfnamefont{D.}~\bibnamefont{Hilton}}, \bibinfo {author}
  {\bibfnamefont{R.}~\bibnamefont{Prasankumar}}, \bibinfo {author}
  {\bibfnamefont{S.}~\bibnamefont{Fourmaux}}, \bibinfo {author}
  {\bibfnamefont{A.}~\bibnamefont{Cavalleri}}, \bibinfo {author}
  {\bibfnamefont{D.}~\bibnamefont{Brassard}}, \bibinfo {author}
  {\bibfnamefont{M.}~\bibnamefont{{El Khakani}}}, \bibinfo {author}
  {\bibfnamefont{J.}~\bibnamefont{Kieffer}}, \bibinfo {author}
  {\bibfnamefont{A.}~\bibnamefont{Taylor}},\ and\ \bibinfo {author}
  {\bibfnamefont{R.}~\bibnamefont{Averitt}},\ }%
  \bibfield{journal}{%
  \Doi{10.1103/PhysRevLett.99.226401}{\bibinfo {journal} {Physical Review
  Letters}}\ }%
  \textbf{\bibinfo {volume} {99}},\ \bibinfo {pages} {226401} (\bibinfo {month}
  {Nov.}\ \bibinfo {year} {2007}),\ ISSN \bibinfo {issn} {0031-9007},\
  \url{http://link.aps.org/doi/10.1103/PhysRevLett.99.226401}%
  \bibAnnoteFile{NoStop}{Hilton2007}%
\bibitem{Muraoka2002}%
  \BibitemOpen
  \bibfield{author}{%
  \bibinfo {author} {\bibfnamefont{Y.}~\bibnamefont{Muraoka}}, \bibinfo
  {author} {\bibfnamefont{Y.}~\bibnamefont{Ueda}},\ and\ \bibinfo {author}
  {\bibfnamefont{Z.}~\bibnamefont{Hiroi}},\ }%
  \bibfield{journal}{%
  \Doi{10.1016/S0022-3697(02)00098-7}{\bibinfo {journal} {Journal of Physics
  and Chemistry of Solids}}\ }%
  \textbf{\bibinfo {volume} {63}},\ \bibinfo {pages} {965} (\bibinfo {month}
  {Aug.}\ \bibinfo {year} {2002}),\ ISSN \bibinfo {issn} {00223697},\
  \url{http://linkinghub.elsevier.com/retrieve/pii/S0022369702000987}%
  \bibAnnoteFile{NoStop}{Muraoka2002}%
\bibitem{Lu2008}%
  \BibitemOpen
  \bibfield{author}{%
  \bibinfo {author} {\bibfnamefont{J.}~\bibnamefont{Lu}}, \bibinfo {author}
  {\bibfnamefont{K.~G.}\ \bibnamefont{West}},\ and\ \bibinfo {author}
  {\bibfnamefont{S.~a.}\ \bibnamefont{Wolf}},\ }%
  \bibfield{journal}{%
  \Doi{10.1063/1.3058769}{\bibinfo {journal} {Applied Physics Letters}}\ }%
  \textbf{\bibinfo {volume} {93}},\ \bibinfo {pages} {262107} (\bibinfo {year}
  {2008}),\ ISSN \bibinfo {issn} {00036951},\
  \url{http://link.aip.org/link/APPLAB/v93/i26/p262107/s1\&Agg=doi}%
  \bibAnnoteFile{NoStop}{Lu2008}%
\bibitem{Zhang2009}%
  \BibitemOpen
  \bibfield{author}{%
  \bibinfo {author} {\bibfnamefont{S.}~\bibnamefont{Zhang}}, \bibinfo {author}
  {\bibfnamefont{J.~Y.}\ \bibnamefont{Chou}},\ and\ \bibinfo {author}
  {\bibfnamefont{L.~J.}\ \bibnamefont{Lauhon}},\ }%
  \bibfield{journal}{%
  \Doi{10.1021/nl9028973}{\bibinfo {journal} {Nano Letters}}\ }%
  \textbf{\bibinfo {volume} {9}},\ \bibinfo {pages} {4527} (\bibinfo {month}
  {Dec.}\ \bibinfo {year} {2009}),\ ISSN \bibinfo {issn} {1530-6992},\
  \url{http://www.ncbi.nlm.nih.gov/pubmed/19902918}%
  \bibAnnoteFile{NoStop}{Zhang2009}%
\bibitem{West2008}%
  \BibitemOpen
  \bibfield{author}{%
  \bibinfo {author} {\bibfnamefont{K.~G.}\ \bibnamefont{West}}, \bibinfo
  {author} {\bibfnamefont{J.}~\bibnamefont{Lu}}, \bibinfo {author}
  {\bibfnamefont{J.}~\bibnamefont{Yu}}, \bibinfo {author}
  {\bibfnamefont{D.}~\bibnamefont{Kirkwood}}, \bibinfo {author}
  {\bibfnamefont{W.}~\bibnamefont{Chen}}, \bibinfo {author}
  {\bibfnamefont{Y.}~\bibnamefont{Pei}}, \bibinfo {author}
  {\bibfnamefont{J.}~\bibnamefont{Claassen}},\ and\ \bibinfo {author}
  {\bibfnamefont{S.~a.}\ \bibnamefont{Wolf}},\ }%
  \bibfield{journal}{%
  \Doi{10.1116/1.2819268}{\bibinfo {journal} {Journal of Vacuum Science \&
  Technology A}}\ }%
  \textbf{\bibinfo {volume} {26}},\ \bibinfo {pages} {133} (\bibinfo {year}
  {2008}),\ ISSN \bibinfo {issn} {07342101},\
  \url{http://link.aip.org/link/JVTAD6/v26/i1/p133/s1\&Agg=doi}%
  \bibAnnoteFile{NoStop}{West2008}%
\bibitem{Jepsen2011}%
  \BibitemOpen
  \bibfield{author}{%
  \bibinfo {author} {\bibfnamefont{P.}~\bibnamefont{Jepsen}}, \bibinfo {author}
  {\bibfnamefont{D.}~\bibnamefont{Cooke}},\ and\ \bibinfo {author}
  {\bibfnamefont{M.}~\bibnamefont{Koch}},\ }%
  \bibfield{journal}{%
  \Doi{10.1002/lpor.201000011}{\bibinfo {journal} {Laser \& Photonics
  Reviews}}\ }%
  \textbf{\bibinfo {volume} {5}},\ \bibinfo {pages} {124} (\bibinfo {month}
  {Jan.}\ \bibinfo {year} {2011}),\ ISSN \bibinfo {issn} {18638880},\
  \url{http://doi.wiley.com/10.1002/lpor.201000011}%
  \bibAnnoteFile{NoStop}{Jepsen2011}%
\bibitem{Jordens2009}%
  \BibitemOpen
  \bibfield{author}{%
  \bibinfo {author} {\bibfnamefont{C.}~\bibnamefont{J\"{o}rdens}}, \bibinfo
  {author} {\bibfnamefont{M.}~\bibnamefont{Scheller}}, \bibinfo {author}
  {\bibfnamefont{M.}~\bibnamefont{Wichmann}}, \bibinfo {author}
  {\bibfnamefont{M.}~\bibnamefont{Mikulics}}, \bibinfo {author}
  {\bibfnamefont{K.}~\bibnamefont{Wiesauer}},\ and\ \bibinfo {author}
  {\bibfnamefont{M.}~\bibnamefont{Koch}},\ }%
  \bibfield{journal}{%
  \bibinfo {journal} {Applied optics}\ }%
  \textbf{\bibinfo {volume} {48}},\ \bibinfo {pages} {2037} (\bibinfo {month}
  {Apr.}\ \bibinfo {year} {2009}),\ ISSN \bibinfo {issn} {1539-4522},\
  \url{http://www.ncbi.nlm.nih.gov/pubmed/19363540}%
  \bibAnnoteFile{NoStop}{Jordens2009}%
\bibitem{Cao2009}%
  \BibitemOpen
  \bibfield{author}{%
  \bibinfo {author} {\bibfnamefont{J.}~\bibnamefont{Cao}}, \bibinfo {author}
  {\bibfnamefont{E.}~\bibnamefont{Ertekin}}, \bibinfo {author}
  {\bibfnamefont{V.}~\bibnamefont{Srinivasan}}, \bibinfo {author}
  {\bibfnamefont{W.}~\bibnamefont{Fan}}, \bibinfo {author}
  {\bibfnamefont{S.}~\bibnamefont{Huang}}, \bibinfo {author}
  {\bibfnamefont{H.}~\bibnamefont{Zheng}}, \bibinfo {author}
  {\bibfnamefont{J.~W.~L.}\ \bibnamefont{Yim}}, \bibinfo {author}
  {\bibfnamefont{D.~R.}\ \bibnamefont{Khanal}}, \bibinfo {author}
  {\bibfnamefont{D.~F.}\ \bibnamefont{Ogletree}}, \bibinfo {author}
  {\bibfnamefont{J.~C.}\ \bibnamefont{Grossman}},\ and\ \bibinfo {author}
  {\bibfnamefont{J.}~\bibnamefont{Wu}},\ }%
  \bibfield{journal}{%
  \Doi{10.1038/nnano.2009.266}{\bibinfo {journal} {Nature Nanotechnology}}\ }%
  \textbf{\bibinfo {volume} {4}},\ \bibinfo {pages} {732} (\bibinfo {month}
  {Sep.}\ \bibinfo {year} {2009}),\ ISSN \bibinfo {issn} {1748-3387},\
  \url{http://www.nature.com/doifinder/10.1038/nnano.2009.266}%
  \bibAnnoteFile{NoStop}{Cao2009}%
\bibitem{Wu2006}%
  \BibitemOpen
  \bibfield{author}{%
  \bibinfo {author} {\bibfnamefont{J.}~\bibnamefont{Wu}}, \bibinfo {author}
  {\bibfnamefont{Q.}~\bibnamefont{Gu}}, \bibinfo {author}
  {\bibfnamefont{B.~S.}\ \bibnamefont{Guiton}}, \bibinfo {author}
  {\bibfnamefont{N.~P.}\ \bibnamefont{de~Leon}}, \bibinfo {author}
  {\bibfnamefont{L.}~\bibnamefont{Ouyang}},\ and\ \bibinfo {author}
  {\bibfnamefont{H.}~\bibnamefont{Park}},\ }%
  \bibfield{journal}{%
  \Doi{10.1021/nl061831r}{\bibinfo {journal} {Nano Letters}}\ }%
  \textbf{\bibinfo {volume} {6}},\ \bibinfo {pages} {2313} (\bibinfo {month}
  {Oct.}\ \bibinfo {year} {2006}),\ ISSN \bibinfo {issn} {1530-6984},\
  \url{http://www.ncbi.nlm.nih.gov/pubmed/17034103}%
  \bibAnnoteFile{NoStop}{Wu2006}%
\bibitem{Sohn2009}%
  \BibitemOpen
  \bibfield{author}{%
  \bibinfo {author} {\bibfnamefont{J.~I.}\ \bibnamefont{Sohn}}, \bibinfo
  {author} {\bibfnamefont{H.~J.}\ \bibnamefont{Joo}}, \bibinfo {author}
  {\bibfnamefont{D.}~\bibnamefont{Ahn}}, \bibinfo {author}
  {\bibfnamefont{H.~H.}\ \bibnamefont{Lee}}, \bibinfo {author}
  {\bibfnamefont{A.~E.}\ \bibnamefont{Porter}}, \bibinfo {author}
  {\bibfnamefont{K.}~\bibnamefont{Kim}}, \bibinfo {author}
  {\bibfnamefont{D.~J.}\ \bibnamefont{Kang}},\ and\ \bibinfo {author}
  {\bibfnamefont{M.~E.}\ \bibnamefont{Welland}},\ }%
  \bibfield{journal}{%
  \Doi{10.1021/nl900841k}{\bibinfo {journal} {Nano Letters}}\ }%
  \textbf{\bibinfo {volume} {9}},\ \bibinfo {pages} {3392} (\bibinfo {month}
  {Oct.}\ \bibinfo {year} {2009}),\ ISSN \bibinfo {issn} {1530-6992},\
  \url{http://www.ncbi.nlm.nih.gov/pubmed/19785429}%
  \bibAnnoteFile{NoStop}{Sohn2009}%
\bibitem{Jones2010}%
  \BibitemOpen
  \bibfield{author}{%
  \bibinfo {author} {\bibfnamefont{A.~C.}\ \bibnamefont{Jones}}, \bibinfo
  {author} {\bibfnamefont{S.}~\bibnamefont{Berweger}}, \bibinfo {author}
  {\bibfnamefont{J.}~\bibnamefont{Wei}}, \bibinfo {author}
  {\bibfnamefont{D.}~\bibnamefont{Cobden}},\ and\ \bibinfo {author}
  {\bibfnamefont{M.~B.}\ \bibnamefont{Raschke}},\ }%
  \bibfield{journal}{%
  \Doi{10.1021/nl903765h}{\bibinfo {journal} {Nano Letters}}\ }%
  \textbf{\bibinfo {volume} {10}},\ \bibinfo {pages} {1574} (\bibinfo {month}
  {May}\ \bibinfo {year} {2010}),\ ISSN \bibinfo {issn} {1530-6992},\
  \url{http://www.ncbi.nlm.nih.gov/pubmed/20377237}%
  \bibAnnoteFile{NoStop}{Jones2010}%
\bibitem{McWhan1974}%
  \BibitemOpen
  \bibfield{author}{%
  \bibinfo {author} {\bibfnamefont{D.~B.}\ \bibnamefont{McWhan}}, \bibinfo
  {author} {\bibfnamefont{M.}~\bibnamefont{Marezio}}, \bibinfo {author}
  {\bibfnamefont{J.~P.}\ \bibnamefont{Remeika}},\ and\ \bibinfo {author}
  {\bibfnamefont{P.~D.}\ \bibnamefont{Dernier}},\ }%
  \bibfield{journal}{%
  \bibinfo {journal} {Physical Review B}\ }%
  \textbf{\bibinfo {volume} {10}},\ \bibinfo {pages} {490} (\bibinfo {year}
  {1974})%
  \bibAnnoteFile{NoStop}{McWhan1974}%
\bibitem{Kucharczyk1979}%
  \BibitemOpen
  \bibfield{author}{%
  \bibinfo {author} {\bibfnamefont{D.}~\bibnamefont{Kucharczyk}}\ and\ \bibinfo
  {author} {\bibfnamefont{T.}~\bibnamefont{Niklewski}},\ }%
  \bibfield{journal}{%
  \bibinfo {journal} {Journal of Applied Crystallography}\ }%
  \textbf{\bibinfo {volume} {12}},\ \bibinfo {pages} {370} (\bibinfo {year}
  {1979})%
  \bibAnnoteFile{NoStop}{Kucharczyk1979}%
\bibitem{Goodenough1971}%
  \BibitemOpen
  \bibfield{author}{%
  \bibinfo {author} {\bibfnamefont{J.}~\bibnamefont{Goodenough}},\ }%
  \bibfield{journal}{%
  \Doi{10.1016/0022-4596(71)90091-0}{\bibinfo {journal} {Journal of Solid State
  Chemistry}}\ }%
  \textbf{\bibinfo {volume} {3}},\ \bibinfo {pages} {490} (\bibinfo {month}
  {Nov.}\ \bibinfo {year} {1971}),\ ISSN \bibinfo {issn} {00224596},\
  \url{http://linkinghub.elsevier.com/retrieve/pii/0022459671900910}%
  \bibAnnoteFile{NoStop}{Goodenough1971}%
\bibitem{Eyert2002}%
  \BibitemOpen
  \bibfield{author}{%
  \bibinfo {author} {\bibfnamefont{V.}~\bibnamefont{Eyert}},\ }%
  \bibfield{journal}{%
  \Doi{10.1002/1521-3889(200210)11:9<650::AID-ANDP650>3.0.CO;2-K}{\bibinfo
  {journal} {Annalen der Physik (Leipzig)}}\ }%
  \textbf{\bibinfo {volume} {11}},\ \bibinfo {pages} {650} (\bibinfo {month}
  {Oct.}\ \bibinfo {year} {2002}),\ ISSN \bibinfo {issn} {00033804},\
  \url{http://doi.wiley.com/10.1002/1521-3889(200210)11:9<650::AID-ANDP650>3.0%
.CO;2-K}%
  \bibAnnoteFile{NoStop}{Eyert2002}%
\bibitem{Zylbersztejn1975}%
  \BibitemOpen
  \bibfield{author}{%
  \bibinfo {author} {\bibfnamefont{A.}~\bibnamefont{Zylbersztejn}}\ and\
  \bibinfo {author} {\bibfnamefont{M.~N.}\ \bibnamefont{F.}},\ }%
  \bibfield{journal}{%
  \bibinfo {journal} {Physical Review B}\ }%
  \textbf{\bibinfo {volume} {11}},\ \bibinfo {pages} {4383} (\bibinfo {year}
  {1975})%
  \bibAnnoteFile{NoStop}{Zylbersztejn1975}%
\bibitem{Berglund1969}%
  \BibitemOpen
  \bibfield{author}{%
  \bibinfo {author} {\bibfnamefont{C.~N.}\ \bibnamefont{Berglund}}\ and\
  \bibinfo {author} {\bibfnamefont{H.~J.}\ \bibnamefont{Guggenheim}},\ }%
  \bibfield{journal}{%
  \bibinfo {journal} {Physical Review}\ }%
  \textbf{\bibinfo {volume} {185}},\ \bibinfo {pages} {1022} (\bibinfo {year}
  {1969})%
  \bibAnnoteFile{NoStop}{Berglund1969}%
\bibitem{Nagashima2006}%
  \BibitemOpen
  \bibfield{author}{%
  \bibinfo {author} {\bibfnamefont{K.}~\bibnamefont{Nagashima}}, \bibinfo
  {author} {\bibfnamefont{T.}~\bibnamefont{Yanagida}}, \bibinfo {author}
  {\bibfnamefont{H.}~\bibnamefont{Tanaka}},\ and\ \bibinfo {author}
  {\bibfnamefont{T.}~\bibnamefont{Kawai}},\ }%
  \bibfield{journal}{%
  \Doi{10.1103/PhysRevB.74.172106}{\bibinfo {journal} {Physical Review B}}\ }%
  \textbf{\bibinfo {volume} {74}},\ \bibinfo {pages} {172106} (\bibinfo {month}
  {Nov.}\ \bibinfo {year} {2006}),\ ISSN \bibinfo {issn} {1098-0121},\
  \url{http://link.aps.org/doi/10.1103/PhysRevB.74.172106}%
  \bibAnnoteFile{NoStop}{Nagashima2006}%
\bibitem{Qazilbash2006}%
  \BibitemOpen
  \bibfield{author}{%
  \bibinfo {author} {\bibfnamefont{M.}~\bibnamefont{Qazilbash}}, \bibinfo
  {author} {\bibfnamefont{K.}~\bibnamefont{Burch}}, \bibinfo {author}
  {\bibfnamefont{D.}~\bibnamefont{Whisler}}, \bibinfo {author}
  {\bibfnamefont{D.}~\bibnamefont{Shrekenhamer}}, \bibinfo {author}
  {\bibfnamefont{B.}~\bibnamefont{Chae}}, \bibinfo {author}
  {\bibfnamefont{H.}~\bibnamefont{Kim}},\ and\ \bibinfo {author}
  {\bibfnamefont{D.}~\bibnamefont{Basov}},\ }%
  \bibfield{journal}{%
  \Doi{10.1103/PhysRevB.74.205118}{\bibinfo {journal} {Physical Review B}}\ }%
  \textbf{\bibinfo {volume} {74}},\ \bibinfo {pages} {205118} (\bibinfo {month}
  {Nov.}\ \bibinfo {year} {2006}),\ ISSN \bibinfo {issn} {1098-0121},\
  \url{http://link.aps.org/doi/10.1103/PhysRevB.74.205118}%
  \bibAnnoteFile{NoStop}{Qazilbash2006}%
\bibitem{Pouget1975}%
  \BibitemOpen
  \bibfield{author}{%
  \bibinfo {author} {\bibfnamefont{J.~P.}\ \bibnamefont{Pouget}}, \bibinfo
  {author} {\bibfnamefont{H.}~\bibnamefont{Launois}}, \bibinfo {author}
  {\bibfnamefont{J.~P.}\ \bibnamefont{D'Haenens}}, \bibinfo {author}
  {\bibfnamefont{P.}~\bibnamefont{Merenda}},\ and\ \bibinfo {author}
  {\bibfnamefont{T.~M.}\ \bibnamefont{Rice}},\ }%
  \bibfield{journal}{%
  \bibinfo {journal} {Physical Review Letters}\ }%
  \textbf{\bibinfo {volume} {35}},\ \bibinfo {pages} {873} (\bibinfo {year}
  {1975})%
  \bibAnnoteFile{NoStop}{Pouget1975}%
\bibitem{Goodenough1973}%
  \BibitemOpen
  \bibfield{author}{%
  \bibinfo {author} {\bibfnamefont{J.~B.}\ \bibnamefont{Goodenough}}\ and\
  \bibinfo {author} {\bibfnamefont{H.~Y.-P.}\ \bibnamefont{Hong}},\ }%
  \bibfield{journal}{%
  \bibinfo {journal} {Physical Review B}\ }%
  \textbf{\bibinfo {volume} {8}},\ \bibinfo {pages} {1323} (\bibinfo {year}
  {1973})%
  \bibAnnoteFile{NoStop}{Goodenough1973}%
\bibitem{Marezio1972}%
  \BibitemOpen
  \bibfield{author}{%
  \bibinfo {author} {\bibfnamefont{M.}~\bibnamefont{Marezio}}, \bibinfo
  {author} {\bibfnamefont{D.~B.}\ \bibnamefont{McWhan}}, \bibinfo {author}
  {\bibfnamefont{J.~P.}\ \bibnamefont{Remeika}},\ and\ \bibinfo {author}
  {\bibfnamefont{P.~D.}\ \bibnamefont{Dernier}},\ }%
  \bibfield{journal}{%
  \bibinfo {journal} {Physical Review B}\ }%
  \textbf{\bibinfo {volume} {5}},\ \bibinfo {pages} {2541} (\bibinfo {year}
  {1972})%
  \bibAnnoteFile{NoStop}{Marezio1972}%
\bibitem{Pouget1976}%
  \BibitemOpen
  \bibfield{author}{%
  \bibinfo {author} {\bibfnamefont{J.~P.}\ \bibnamefont{Pouget}}\ and\ \bibinfo
  {author} {\bibfnamefont{H.}~\bibnamefont{Launois}},\ }%
  \bibfield{journal}{%
  \bibinfo {journal} {Journal de Physique Colloques}\ }%
  \textbf{\bibinfo {volume} {37}},\ \bibinfo {pages} {C4} (\bibinfo {year}
  {1976})%
  \bibAnnoteFile{NoStop}{Pouget1976}%
\bibitem{Villeneuve1973}%
  \BibitemOpen
  \bibfield{author}{%
  \bibinfo {author} {\bibfnamefont{P.}~\bibnamefont{{Villeneuve, G., Drillon,
  M., Hagenmuller}}},\ }%
  \bibfield{journal}{%
  \bibinfo {journal} {Materials Research Bulletin}\ }%
  \textbf{\bibinfo {volume} {8}},\ \bibinfo {pages} {1111} (\bibinfo {year}
  {1973})%
  \bibAnnoteFile{NoStop}{Villeneuve1973}%
\bibitem{MacChesney1969}%
  \BibitemOpen
  \bibfield{author}{%
  \bibinfo {author} {\bibfnamefont{J.}~\bibnamefont{Macchesney}}\ and\ \bibinfo
  {author} {\bibfnamefont{H.}~\bibnamefont{Guggenheim}},\ }%
  \bibfield{journal}{%
  \Doi{10.1016/0022-3697(69)90303-5}{\bibinfo {journal} {Journal of Physics and
  Chemistry of Solids}}\ }%
  \textbf{\bibinfo {volume} {30}},\ \bibinfo {pages} {225} (\bibinfo {month}
  {Feb.}\ \bibinfo {year} {1969}),\ ISSN \bibinfo {issn} {00223697},\
  \url{http://linkinghub.elsevier.com/retrieve/pii/0022369769903035}%
  \bibAnnoteFile{NoStop}{MacChesney1969}%
\bibitem{McWhan1969}%
  \BibitemOpen
  \bibfield{author}{%
  \bibinfo {author} {\bibfnamefont{D.~B.}\ \bibnamefont{McWhan}}, \bibinfo
  {author} {\bibfnamefont{T.~M.}\ \bibnamefont{Rice}},\ and\ \bibinfo {author}
  {\bibfnamefont{J.~P.}\ \bibnamefont{Remeika}},\ }%
  \bibfield{journal}{%
  \bibinfo {journal} {Physical Review Letters}\ }%
  \textbf{\bibinfo {volume} {23}},\ \bibinfo {pages} {1384} (\bibinfo {year}
  {1969})%
  \bibAnnoteFile{NoStop}{McWhan1969}%
\bibitem{Barker1966}%
  \BibitemOpen
  \bibfield{author}{%
  \bibinfo {author} {\bibfnamefont{H.~J.}\ \bibnamefont{{Barker, A. S.,
  Verleur, H. W., Guggenheim}}},\ }%
  \bibfield{journal}{%
  \bibinfo {journal} {Physical Review Letters}\ }%
  \textbf{\bibinfo {volume} {17}},\ \bibinfo {pages} {1286} (\bibinfo {year}
  {1966})%
  \bibAnnoteFile{NoStop}{Barker1966}%
\bibitem{Everhart1968}%
  \BibitemOpen
  \bibfield{author}{%
  \bibinfo {author} {\bibfnamefont{J.~B.}\ \bibnamefont{{Everhart, C. R.,
  MacChesney}}},\ }%
  \bibfield{journal}{%
  \bibinfo {journal} {Journal of Applied Physics}\ }%
  \textbf{\bibinfo {volume} {39}},\ \bibinfo {pages} {2872} (\bibinfo {year}
  {1968})%
  \bibAnnoteFile{NoStop}{Everhart1968}%
\bibitem{Bongers1965}%
  \BibitemOpen
  \bibfield{author}{%
  \bibinfo {author} {\bibfnamefont{P.~F.}\ \bibnamefont{Bongers}},\ }%
  \bibfield{journal}{%
  \bibinfo {journal} {Solid State Communications}\ }%
  \textbf{\bibinfo {volume} {3}},\ \bibinfo {pages} {275} (\bibinfo {year}
  {1965})%
  \bibAnnoteFile{NoStop}{Bongers1965}%
\bibitem{Kosuge1967}%
  \BibitemOpen
  \bibfield{author}{%
  \bibinfo {author} {\bibfnamefont{K.}~\bibnamefont{Kosuge}},\ }%
  \bibfield{journal}{%
  \bibinfo {journal} {Journal of the Physical Society of Japan}\ }%
  \textbf{\bibinfo {volume} {22}},\ \bibinfo {pages} {551} (\bibinfo {year}
  {1967})%
  \bibAnnoteFile{NoStop}{Kosuge1967}%
\bibitem{Neuman1964}%
  \BibitemOpen
  \bibfield{author}{%
  \bibinfo {author} {\bibfnamefont{C.~H.}\ \bibnamefont{Neuman}}, \bibinfo
  {author} {\bibfnamefont{A.~W.}\ \bibnamefont{Lawson}},\ and\ \bibinfo
  {author} {\bibfnamefont{R.~F.}\ \bibnamefont{Brown}},\ }%
  \bibfield{journal}{%
  \bibinfo {journal} {The Journal of Chemical Physics}\ }%
  \textbf{\bibinfo {volume} {41}},\ \bibinfo {pages} {1591} (\bibinfo {year}
  {1964})%
  \bibAnnoteFile{NoStop}{Neuman1964}%
\bibitem{Berglund1969Hyd}%
  \BibitemOpen
  \bibfield{author}{%
  \bibinfo {author} {\bibfnamefont{C.~N.}\ \bibnamefont{Berglund}}\ and\
  \bibinfo {author} {\bibfnamefont{A.}~\bibnamefont{Jayamaran}},\ }%
  \bibfield{journal}{%
  \bibinfo {journal} {Physical Review}\ }%
  \textbf{\bibinfo {volume} {185}},\ \bibinfo {pages} {1034} (\bibinfo {year}
  {1969})%
  \bibAnnoteFile{NoStop}{Berglund1969Hyd}%
\bibitem{Sepulveda2008}%
  \BibitemOpen
  \bibfield{author}{%
  \bibinfo {author} {\bibfnamefont{N.}~\bibnamefont{Sepúlveda}}, \bibinfo
  {author} {\bibfnamefont{A.}~\bibnamefont{Rúa}}, \bibinfo {author}
  {\bibfnamefont{R.}~\bibnamefont{Cabrera}},\ and\ \bibinfo {author}
  {\bibfnamefont{F.}~\bibnamefont{Fernández}},\ }%
  \bibfield{journal}{%
  \Doi{10.1063/1.2926681}{\bibinfo {journal} {Applied Physics Letters}}\ }%
  \textbf{\bibinfo {volume} {92}},\ \bibinfo {pages} {191913} (\bibinfo {year}
  {2008}),\ ISSN \bibinfo {issn} {00036951},\
  \url{http://link.aip.org/link/APPLAB/v92/i19/p191913/s1\&Agg=doi}%
  \bibAnnoteFile{NoStop}{Sepulveda2008}%
\bibitem{Beteille1998}%
  \BibitemOpen
  \bibfield{author}{%
  \bibinfo {author} {\bibfnamefont{F.}~\bibnamefont{B\'{e}teille}}\ and\
  \bibinfo {author} {\bibfnamefont{J.}~\bibnamefont{Livage}},\ }%
  \bibfield{journal}{%
  \bibinfo {journal} {Journal of Sol-Gel Science and Technology}\ }%
  \textbf{\bibinfo {volume} {13}},\ \bibinfo {pages} {915} (\bibinfo {year}
  {1998})%
  \bibAnnoteFile{NoStop}{Beteille1998}%
\bibitem{Villeneuve1972}%
  \BibitemOpen
  \bibfield{author}{%
  \bibinfo {author} {\bibfnamefont{G.}~\bibnamefont{Villeneuve}}, \bibinfo
  {author} {\bibfnamefont{A.}~\bibnamefont{Bordet}}, \bibinfo {author}
  {\bibfnamefont{A.}~\bibnamefont{Casalot}}, \bibinfo {author}
  {\bibfnamefont{J.}~\bibnamefont{Pouget}}, \bibinfo {author}
  {\bibfnamefont{H.}~\bibnamefont{Launois}},\ and\ \bibinfo {author}
  {\bibfnamefont{P.}~\bibnamefont{Lederer}},\ }%
  \bibfield{journal}{%
  \Doi{10.1016/0038-1098(72)91191-X}{\bibinfo {journal} {Journal of Physics and
  Chemistry of Solids}}\ }%
  \textbf{\bibinfo {volume} {33}},\ \bibinfo {pages} {1953} (\bibinfo {month}
  {Jul.}\ \bibinfo {year} {1972}),\ ISSN \bibinfo {issn} {00381098},\
  \url{http://linkinghub.elsevier.com/retrieve/pii/003810987291191X}%
  \bibAnnoteFile{NoStop}{Villeneuve1972}%
\bibitem{Piccirillo2007}%
  \BibitemOpen
  \bibfield{author}{%
  \bibinfo {author} {\bibfnamefont{C.}~\bibnamefont{Piccirillo}}, \bibinfo
  {author} {\bibfnamefont{R.}~\bibnamefont{Binions}},\ and\ \bibinfo {author}
  {\bibfnamefont{I.~P.}\ \bibnamefont{Parkin}},\ }%
  \bibfield{journal}{%
  \Doi{10.1002/ejic.200700284}{\bibinfo {journal} {European Journal of
  Inorganic Chemistry}}\ }%
  \textbf{\bibinfo {volume} {2007}},\ \bibinfo {pages} {4050} (\bibinfo {month}
  {Sep.}\ \bibinfo {year} {2007}),\ ISSN \bibinfo {issn} {14341948},\
  \url{http://doi.wiley.com/10.1002/ejic.200700284}%
  \bibAnnoteFile{NoStop}{Piccirillo2007}%
\bibitem{Horlin1973}%
  \BibitemOpen
  \bibfield{author}{%
  \bibinfo {author} {\bibfnamefont{M.}~\bibnamefont{{H\"{o}rlin, T., Niklewski,
  T., Nygren}}},\ }%
  \bibfield{journal}{%
  \bibinfo {journal} {Materials Research Bulletin}\ }%
  \textbf{\bibinfo {volume} {8}},\ \bibinfo {pages} {179} (\bibinfo {year}
  {1973})%
  \bibAnnoteFile{NoStop}{Horlin1973}%
\bibitem{Holman2009}%
  \BibitemOpen
  \bibfield{author}{%
  \bibinfo {author} {\bibfnamefont{K.}~\bibnamefont{Holman}}, \bibinfo {author}
  {\bibfnamefont{T.}~\bibnamefont{McQueen}}, \bibinfo {author}
  {\bibfnamefont{A.}~\bibnamefont{Williams}}, \bibinfo {author}
  {\bibfnamefont{T.}~\bibnamefont{Klimczuk}}, \bibinfo {author}
  {\bibfnamefont{P.}~\bibnamefont{Stephens}}, \bibinfo {author}
  {\bibfnamefont{H.}~\bibnamefont{Zandbergen}}, \bibinfo {author}
  {\bibfnamefont{Q.}~\bibnamefont{Xu}}, \bibinfo {author}
  {\bibfnamefont{F.}~\bibnamefont{Ronning}},\ and\ \bibinfo {author}
  {\bibfnamefont{R.}~\bibnamefont{Cava}},\ }%
  \bibfield{journal}{%
  \Doi{10.1103/PhysRevB.79.245114}{\bibinfo {journal} {Physical Review B}}\ }%
  \textbf{\bibinfo {volume} {79}},\ \bibinfo {pages} {245114} (\bibinfo {month}
  {Jun.}\ \bibinfo {year} {2009}),\ ISSN \bibinfo {issn} {1098-0121},\
  \url{http://link.aps.org/doi/10.1103/PhysRevB.79.245114}%
  \bibAnnoteFile{NoStop}{Holman2009}%
\bibitem{Nygren1969}%
  \BibitemOpen
  \bibfield{author}{%
  \bibinfo {author} {\bibfnamefont{M.}~\bibnamefont{{Nygren, M.,
  Israelsson}}},\ }%
  \bibfield{journal}{%
  \bibinfo {journal} {Materials Research Bulletin}\ }%
  \textbf{\bibinfo {volume} {4}},\ \bibinfo {pages} {881} (\bibinfo {year}
  {1969})%
  \bibAnnoteFile{NoStop}{Nygren1969}%
\bibitem{Chae2008}%
  \BibitemOpen
  \bibfield{author}{%
  \bibinfo {author} {\bibfnamefont{B.~G.}\ \bibnamefont{Chae}}, \bibinfo
  {author} {\bibfnamefont{H.~T.}\ \bibnamefont{Kim}},\ and\ \bibinfo {author}
  {\bibfnamefont{S.~J.}\ \bibnamefont{Yun}},\ }%
  \bibfield{journal}{%
  \Doi{10.1149/1.2903208}{\bibinfo {journal} {Electrochemical and Solid-State
  Letters}}\ }%
  \textbf{\bibinfo {volume} {11}},\ \bibinfo {pages} {D53} (\bibinfo {year}
  {2008}),\ ISSN \bibinfo {issn} {10990062},\
  \url{http://link.aip.org/link/ESLEF6/v11/i6/pD53/s1\&Agg=doi}%
  \bibAnnoteFile{NoStop}{Chae2008}%
\bibitem{Savborg1977}%
  \BibitemOpen
  \bibfield{author}{%
  \bibinfo {author} {\bibfnamefont{O.}~\bibnamefont{S\"{a}vborg}}\ and\
  \bibinfo {author} {\bibfnamefont{M.}~\bibnamefont{Nygren}},\ }%
  \bibfield{journal}{%
  \Doi{10.1002/pssa.2210430236}{\bibinfo {journal} {Physica Status Solidi
  (a)}}\ }%
  \textbf{\bibinfo {volume} {43}},\ \bibinfo {pages} {645} (\bibinfo {month}
  {Oct.}\ \bibinfo {year} {1977}),\ ISSN \bibinfo {issn} {00318965},\
  \url{http://doi.wiley.com/10.1002/pssa.2210430236}%
  \bibAnnoteFile{NoStop}{Savborg1977}%
\bibitem{Bayard1975}%
  \BibitemOpen
  \bibfield{author}{%
  \bibinfo {author} {\bibfnamefont{M.}~\bibnamefont{Bayard}},\ }%
  \bibfield{journal}{%
  \Doi{10.1016/0022-4596(75)90176-0}{\bibinfo {journal} {Journal of Solid State
  Chemistry}}\ }%
  \textbf{\bibinfo {volume} {12}},\ \bibinfo {pages} {41} (\bibinfo {month}
  {Jan.}\ \bibinfo {year} {1975}),\ ISSN \bibinfo {issn} {00224596},\
  \url{http://linkinghub.elsevier.com/retrieve/pii/0022459675901760}%
  \bibAnnoteFile{NoStop}{Bayard1975}%
\bibitem{Kristensen1968}%
  \BibitemOpen
  \bibfield{author}{%
  \bibinfo {author} {\bibfnamefont{I.~K.}\ \bibnamefont{Kristensen}},\ }%
  \bibfield{journal}{%
  \bibinfo {journal} {Journal of Applied Physics}\ }%
  \textbf{\bibinfo {volume} {39}},\ \bibinfo {pages} {5341} (\bibinfo {year}
  {1968})%
  \bibAnnoteFile{NoStop}{Kristensen1968}%
\bibitem{Marinder1957}%
  \BibitemOpen
  \bibfield{author}{%
  \bibinfo {author} {\bibfnamefont{B.-O.}\ \bibnamefont{Marinder}}\ and\
  \bibinfo {author} {\bibfnamefont{A.}~\bibnamefont{Magn\'{e}li}},\ }%
  \bibfield{journal}{%
  \bibinfo {journal} {Acta Chemica Scandinavica}\ }%
  \textbf{\bibinfo {volume} {11}},\ \bibinfo {pages} {1635} (\bibinfo {year}
  {1957})%
  \bibAnnoteFile{NoStop}{Marinder1957}%
\bibitem{Gu2007}%
  \BibitemOpen
  \bibfield{author}{%
  \bibinfo {author} {\bibfnamefont{Q.}~\bibnamefont{Gu}}, \bibinfo {author}
  {\bibfnamefont{A.}~\bibnamefont{Falk}}, \bibinfo {author}
  {\bibfnamefont{J.}~\bibnamefont{Wu}}, \bibinfo {author}
  {\bibfnamefont{L.}~\bibnamefont{Ouyang}},\ and\ \bibinfo {author}
  {\bibfnamefont{H.}~\bibnamefont{Park}},\ }%
  \bibfield{journal}{%
  \Doi{10.1021/nl0624768}{\bibinfo {journal} {Nano Letters}}\ }%
  \textbf{\bibinfo {volume} {7}},\ \bibinfo {pages} {363} (\bibinfo {month}
  {Mar.}\ \bibinfo {year} {2007}),\ ISSN \bibinfo {issn} {1530-6984},\
  \url{http://www.ncbi.nlm.nih.gov/pubmed/17256915}%
  \bibAnnoteFile{NoStop}{Gu2007}%
\bibitem{Drillon1974}%
  \BibitemOpen
  \bibfield{author}{%
  \bibinfo {author} {\bibfnamefont{M.}~\bibnamefont{Drillon}}\ and\ \bibinfo
  {author} {\bibfnamefont{G.}~\bibnamefont{Villeneuve}},\ }%
  \bibfield{journal}{%
  \bibinfo {journal} {Materials Research Bulletin}\ }%
  \textbf{\bibinfo {volume} {9}},\ \bibinfo {pages} {1199} (\bibinfo {year}
  {1974})%
  \bibAnnoteFile{NoStop}{Drillon1974}%
\bibitem{Pollert1976}%
  \BibitemOpen
  \bibfield{author}{%
  \bibinfo {author} {\bibfnamefont{E.}~\bibnamefont{Pollert}}, \bibinfo
  {author} {\bibfnamefont{G.}~\bibnamefont{Villeneuve}}, \bibinfo {author}
  {\bibfnamefont{F.}~\bibnamefont{M\'{e}nil}},\ and\ \bibinfo {author}
  {\bibfnamefont{P.}~\bibnamefont{Hagenmuller}},\ }%
  \bibfield{journal}{%
  \bibinfo {journal} {Materials Research Bulletin}\ }%
  \textbf{\bibinfo {volume} {11}},\ \bibinfo {pages} {159} (\bibinfo {year}
  {1976})%
  \bibAnnoteFile{NoStop}{Pollert1976}%
\bibitem{Pintchovski1978}%
  \BibitemOpen
  \bibfield{author}{%
  \bibinfo {author} {\bibfnamefont{F.}~\bibnamefont{Pintchovski}}, \bibinfo
  {author} {\bibfnamefont{W.~S.}\ \bibnamefont{Glaunsinger}},\ and\ \bibinfo
  {author} {\bibfnamefont{A.}~\bibnamefont{Navrotsky}},\ }%
  \bibfield{journal}{%
  \bibinfo {journal} {Journal of the Physics and Chemistry of Solids}\ }%
  \textbf{\bibinfo {volume} {39}},\ \bibinfo {pages} {941} (\bibinfo {year}
  {1978})%
  \bibAnnoteFile{NoStop}{Pintchovski1978}%
\bibitem{Kitahiro1967}%
  \BibitemOpen
  \bibfield{author}{%
  \bibinfo {author} {\bibfnamefont{I.}~\bibnamefont{Kitahiro}}\ and\ \bibinfo
  {author} {\bibfnamefont{A.}~\bibnamefont{Watanabe}},\ }%
  \bibfield{journal}{%
  \bibinfo {journal} {Japanese Journal of Applied Physics}\ }%
  \textbf{\bibinfo {volume} {6}},\ \bibinfo {pages} {1023} (\bibinfo {year}
  {1967})%
  \bibAnnoteFile{NoStop}{Kitahiro1967}%
\bibitem{Lee1996}%
  \BibitemOpen
  \bibfield{author}{%
  \bibinfo {author} {\bibfnamefont{M.-H.}\ \bibnamefont{Lee}}, \bibinfo
  {author} {\bibfnamefont{M.-G.}\ \bibnamefont{Kim}},\ and\ \bibinfo {author}
  {\bibfnamefont{H.-K.}\ \bibnamefont{Song}},\ }%
  \bibfield{journal}{%
  \bibinfo {journal} {Thin Solid Films}\ }%
  \textbf{\bibinfo {volume} {290-291}},\ \bibinfo {pages} {30} (\bibinfo {year}
  {1996})%
  \bibAnnoteFile{NoStop}{Lee1996}%
\bibitem{Verleur1968}%
  \BibitemOpen
  \bibfield{author}{%
  \bibinfo {author} {\bibfnamefont{H.~W.}\ \bibnamefont{Verleur}}, \bibinfo
  {author} {\bibfnamefont{A.~S.}\ \bibnamefont{Barker}},\ and\ \bibinfo
  {author} {\bibfnamefont{C.~N.}\ \bibnamefont{Berglund}},\ }%
  \bibfield{journal}{%
  \bibinfo {journal} {Physical Review}\ }%
  \textbf{\bibinfo {volume} {172}},\ \bibinfo {pages} {788} (\bibinfo {year}
  {1968})%
  \bibAnnoteFile{NoStop}{Verleur1968}%
\bibitem{Continenza1999}%
  \BibitemOpen
  \bibfield{author}{%
  \bibinfo {author} {\bibfnamefont{a.}~\bibnamefont{Continenza}}, \bibinfo
  {author} {\bibfnamefont{S.}~\bibnamefont{Massidda}},\ and\ \bibinfo {author}
  {\bibfnamefont{M.}~\bibnamefont{Posternak}},\ }%
  \bibfield{journal}{%
  \Doi{10.1103/PhysRevB.60.15699}{\bibinfo {journal} {Physical Review B}}\ }%
  \textbf{\bibinfo {volume} {60}},\ \bibinfo {pages} {15699} (\bibinfo {month}
  {Dec.}\ \bibinfo {year} {1999}),\ ISSN \bibinfo {issn} {0163-1829},\
  \url{http://link.aps.org/doi/10.1103/PhysRevB.60.15699}%
  \bibAnnoteFile{NoStop}{Continenza1999}%
\bibitem{Mossanek2007}%
  \BibitemOpen
  \bibfield{author}{%
  \bibinfo {author} {\bibfnamefont{R.~J.~O.}\ \bibnamefont{Mossanek}}\ and\
  \bibinfo {author} {\bibfnamefont{M.}~\bibnamefont{Abbate}},\ }%
  \bibfield{journal}{%
  \Doi{10.1088/0953-8984/19/34/346225}{\bibinfo {journal} {Journal of Physics:
  Condensed Matter}}\ }%
  \textbf{\bibinfo {volume} {19}},\ \bibinfo {pages} {346225} (\bibinfo {month}
  {Aug.}\ \bibinfo {year} {2007}),\ ISSN \bibinfo {issn} {0953-8984},\
  \url{http://stacks.iop.org/0953-8984/19/i=34/a=346225?key=crossref.097329a85%
05cf348ca29f0117d3a15db}%
  \bibAnnoteFile{NoStop}{Mossanek2007}%
\bibitem{Lysenko2007}%
  \BibitemOpen
  \bibfield{author}{%
  \bibinfo {author} {\bibfnamefont{S.}~\bibnamefont{Lysenko}}, \bibinfo
  {author} {\bibfnamefont{V.}~\bibnamefont{Vikhnin}}, \bibinfo {author}
  {\bibfnamefont{F.}~\bibnamefont{Fernandez}}, \bibinfo {author}
  {\bibfnamefont{A.}~\bibnamefont{Rua}},\ and\ \bibinfo {author}
  {\bibfnamefont{H.}~\bibnamefont{Liu}},\ }%
  \bibfield{journal}{%
  \Doi{10.1103/PhysRevB.75.075109}{\bibinfo {journal} {Physical Review B}}\ }%
  \textbf{\bibinfo {volume} {75}},\ \bibinfo {pages} {075109} (\bibinfo {month}
  {Feb.}\ \bibinfo {year} {2007}),\ ISSN \bibinfo {issn} {1098-0121},\
  \url{http://link.aps.org/doi/10.1103/PhysRevB.75.075109}%
  \bibAnnoteFile{NoStop}{Lysenko2007}%
\bibitem{Tomczak2009}%
  \BibitemOpen
  \bibfield{author}{%
  \bibinfo {author} {\bibfnamefont{J.}~\bibnamefont{Tomczak}}\ and\ \bibinfo
  {author} {\bibfnamefont{S.}~\bibnamefont{Biermann}},\ }%
  \bibfield{journal}{%
  \Doi{10.1103/PhysRevB.80.085117}{\bibinfo {journal} {Physical Review B}}\ }%
  \textbf{\bibinfo {volume} {80}},\ \bibinfo {pages} {1} (\bibinfo {month}
  {Aug.}\ \bibinfo {year} {2009}),\ ISSN \bibinfo {issn} {1098-0121},\
  \url{http://link.aps.org/doi/10.1103/PhysRevB.80.085117}%
  \bibAnnoteFile{NoStop}{Tomczak2009}%
\bibitem{Minomura1964}%
  \BibitemOpen
  \bibfield{author}{%
  \bibinfo {author} {\bibfnamefont{S.}~\bibnamefont{Minomura}}\ and\ \bibinfo
  {author} {\bibfnamefont{H.}~\bibnamefont{Hagasaki}},\ }%
  \bibfield{journal}{%
  \bibinfo {journal} {Journal of the Physical Society of Japan}\ }%
  \textbf{\bibinfo {volume} {19}},\ \bibinfo {pages} {131} (\bibinfo {year}
  {1964})%
  \bibAnnoteFile{NoStop}{Minomura1964}%
\bibitem{Gregg1997}%
  \BibitemOpen
  \bibfield{author}{%
  \bibinfo {author} {\bibfnamefont{J.~M.}\ \bibnamefont{Gregg}}\ and\ \bibinfo
  {author} {\bibfnamefont{R.~M.}\ \bibnamefont{Bowman}},\ }%
  \bibfield{journal}{%
  \Doi{10.1063/1.120469}{\bibinfo {journal} {Applied Physics Letters}}\ }%
  \textbf{\bibinfo {volume} {71}},\ \bibinfo {pages} {3649} (\bibinfo {year}
  {1997}),\ ISSN \bibinfo {issn} {00036951},\
  \url{http://link.aip.org/link/APPLAB/v71/i25/p3649/s1\&Agg=doi}%
  \bibAnnoteFile{NoStop}{Gregg1997}%
\bibitem{Sakai2008}%
  \BibitemOpen
  \bibfield{author}{%
  \bibinfo {author} {\bibfnamefont{J.}~\bibnamefont{Sakai}},\ }%
  \bibfield{journal}{%
  \Doi{10.1063/1.2986146}{\bibinfo {journal} {Journal of Applied Physics}}\ }%
  \textbf{\bibinfo {volume} {104}},\ \bibinfo {pages} {073703} (\bibinfo {year}
  {2008}),\ ISSN \bibinfo {issn} {00218979},\
  \url{http://link.aip.org/link/JAPIAU/v104/i7/p073703/s1\&Agg=doi}%
  \bibAnnoteFile{NoStop}{Sakai2008}%
\bibitem{Wentzcovitch1994}%
  \BibitemOpen
  \bibfield{author}{%
  \bibinfo {author} {\bibfnamefont{R.~M.}\ \bibnamefont{Wentzcovitch}},
  \bibinfo {author} {\bibfnamefont{W.~W.}\ \bibnamefont{Schulz}},\ and\
  \bibinfo {author} {\bibfnamefont{P.~B.}\ \bibnamefont{Allen}},\ }%
  \bibfield{journal}{%
  \bibinfo {journal} {Physical Review Letters}\ }%
  \textbf{\bibinfo {volume} {72}},\ \bibinfo {pages} {3389} (\bibinfo {year}
  {1994})%
  \bibAnnoteFile{NoStop}{Wentzcovitch1994}%
\bibitem{Tanaka2003}%
  \BibitemOpen
  \bibfield{author}{%
  \bibinfo {author} {\bibfnamefont{A.}~\bibnamefont{Tanaka}},\ }%
  \bibfield{journal}{%
  \Doi{10.1143/JPSJ.72.2433}{\bibinfo {journal} {Journal of the Physics Society
  Japan}}\ }%
  \textbf{\bibinfo {volume} {72}},\ \bibinfo {pages} {2433} (\bibinfo {month}
  {Oct.}\ \bibinfo {year} {2003}),\ ISSN \bibinfo {issn} {0031-9015},\
  \url{http://jpsj.ipap.jp/link?JPSJ/72/2433/}%
  \bibAnnoteFile{NoStop}{Tanaka2003}%
\bibitem{Liebsch2005}%
  \BibitemOpen
  \bibfield{author}{%
  \bibinfo {author} {\bibfnamefont{a.}~\bibnamefont{Liebsch}}, \bibinfo
  {author} {\bibfnamefont{H.}~\bibnamefont{Ishida}},\ and\ \bibinfo {author}
  {\bibfnamefont{G.}~\bibnamefont{Bihlmayer}},\ }%
  \bibfield{journal}{%
  \Doi{10.1103/PhysRevB.71.085109}{\bibinfo {journal} {Physical Review B}}\ }%
  \textbf{\bibinfo {volume} {71}},\ \bibinfo {pages} {085109} (\bibinfo {month}
  {Feb.}\ \bibinfo {year} {2005}),\ ISSN \bibinfo {issn} {1098-0121},\
  \url{http://link.aps.org/doi/10.1103/PhysRevB.71.085109}%
  \bibAnnoteFile{NoStop}{Liebsch2005}%
\end{thebibliography}
\end{document}